\documentclass[12pt,preprint]{aastex}

\shorttitle{Abrupt magnetic changes in flaring regions}
\shortauthors{Petrie \& Sudol}

\begin{document}


\title{Abrupt longitudinal magnetic field changes in flaring
active regions}


\author{G.J.D. Petrie \& J.J. Sudol}
\affil{National Solar Observatory, 950 N. Cherry Avenue, Tucson, AZ 85719\\
West Chester University, West Chester, PA 19383}



\begin{abstract}

\end{abstract}
We characterize the changes in the longitudinal photospheric magnetic field during 38 X-class and 39 M-class flares within $65^{\circ}$ of disk-center using 1-minute GONG magnetograms.  In all 77 cases we identify at least one site in the flaring active region where clear, permanent, stepwise field changes occurred.  The median duration of the field changes was about 15 minutes and was approximately equal for X-class and for M-class flares.  The absolute values of the field changes ranged from the detection limit of $\sim\!\!10$~G to as high as $\sim\!\!450$~G in two exceptional cases.  The median value was 69~G.  Field changes were significantly stronger for X-class than for M-class flares and for limb flares than for disk-center flares.  Longitudinal field changes less than 100~G tended to decrease longitudinal field strengths, both close to disk-center and close to the limb, while field changes greater than 100~G showed no such pattern.  Likewise, longitudinal flux strengths tended to decrease during flares.  Flux changes, particularly net flux changes near disk-center, correlated better than local field changes with GOES peak X-ray flux.  The strongest longitudinal field and flux changes occurred in flares observed close to the limb.  We estimate the change of Lorentz force associated with each flare and find that this is large enough in some cases to power seismic waves.  We find that longitudinal field decreases would likely outnumber increases at all parts of the solar disk within $65^{\circ}$ of disk-center, as in our observations, if photospheric field tilts increase during flares as predicted by Hudson et al.



\section{Introduction}
\label{introduction}

Solar flares are generally believed to be caused by strong, stressed, topologically complicated magnetic fields.  The energy estimated to power a solar flare can only come from the magnetic field and this field must be sufficiently stressed to contain enough free energy to power the flare.  The topology must be complicated enough to contain a magnetic null point for abrupt energy release to be possible (e.g. Priest \& Forbes~2000, Aschwanden~2004).  Magnetic gradient maps derived from GONG longitudinal magnetograms (Gallagher et al.~2002), now available at www.solarmonitor.org, are a useful diagnostic of flare activity (see also Brockman~2010).  Photospheric field gradients have long been known to be related to flare activity in active regions (e.g. Zhang et al.~1994).  For a long time, however, observational studies of flare-related changes in longitudinal (i.e., the component along the observer's line of sight) and vector photospheric magnetic fields were inconclusive because of limitations of instrument sensitivity, spatial resolution, cadence and coverage (Rust~1974, Sakurai \& Hiei~1996).  Moreover, some apparent magnetic field changes associated with large flares were later found not to represent real magnetic field changes but were due to flare-induced changes in the spectral line profiles used in measuring magnetic field strength (Patterson~1984, Harvey 1986, Qiu \& Gary~2003, Edelman et al.~2004).  In the past 15 years or so, however, high-cadence measurements of the photospheric magnetic field have become sensitive enough to resolve fast and permanent field changes in the vicinity of and coincident with large solar flares.

Wang et al.~(1992, 1994) found rapid and permanent field changes in flaring active regions, but a number of later studies produced inconclusive results; see the discussion in Wang~(2006).  Kosovichev \& Zharkova~(1999) reported a sudden decrease in magnetic energy near an X-class flare, during its impulsive phase.  A short time later, Kosovichev \& Zharkova~(2001) reported on regions of permanent decrease of longitudinal magnetic flux in the vicinity of the magnetic neutral line near the 2001 July 14 ``Bastille Day'' flare and linked the change in flux to the release of magnetic energy.  The Big Bear Solar Observatory (BBSO) group has also described numerous cases featuring the sudden appearance and persistence of unbalanced magnetic flux at the time of a flare (Spirock et al.~2002, Wang et al.~2002, 2004, Yurchyshyn et al.~2004).
Using one-minute GONG magnetograms, Sudol \& Harvey~(2005, henceforth referred to as SH05) characterized the spatial distribution, strength and rate of change of permanent field changes associated with 15 X-class flares.  Field changes ranged from 30~G to almost 300~G with a median value of 90~G.  They found that the majority of field changes occured in regions where the field strength reached hundreds of Gauss which suggests locations close to or within sunspots given the resolution of the data.  Liu et al.~(2005) studied one M-class and six X-class flares and reported a roughly even split of increasing and decreasing longitudinal magnetic flux in regions of penumbral decay.

Wang~(2006) studied five flaring $\delta$-spots using high-cadence MDI longitudinal magnetograms and found clear changes in the magnetic gradient along the neutral lines in all cases: the gradient increased in three cases and decreased in two.  The centers of mass of the two magnetic polarities converged/diverged in the cases with gradient increase/decrease.  For 11 data sets where vector data were available, Wang \& Liu~(2010) found that the transverse field at the polarity inversion line invariably increased.  For all but one of 18 cases where 1-minute longitudinal data were available, the limbward flux was observed to increase and the diskward flux to decrease.

The BBSO group has also found a consistent pattern of behavior in sunspot structure.  Parts of the outer penumbral structures decay rapidly after many flares, while neighboring umbral cores and inner penumbal regions become darker (Wang et al.~2004, 2005, Deng et al.~2005, Liu et al.~2005, Wang et al.~2009).  Meanwhile, transverse fields were found to decrease in the regions of penumbral decay and to increase at the flare neutral lines.  Wang et al.~(2002) describe one extreme case in which the onset of an M-class flare coincided with the disappearance of a small sunspot.  Li et al.~(2009) found that during the 2006 December 13 X3.4 flare the mean inclination angle of the magnetic field increased in the part of the penumbra that decayed, whereas the inclination angle decreased in the part of the penumbra that was enhanced during the flare and near the magnetic neutral line.

In contrast, many theoretical models of flares have incorporated the assumption that photospheric fields should not change significantly during flares (e.g., Forbes \& Priest~2002) and this trend continues to the present.  Indeed, Mei \& Lin~(2008) attribute the observed field changes to the fact that the spectral lines used in the observations are not formed in the photosphere.  However, the Ni~{\sc I} line at 676.8 nm used by GONG and MDI corresponds to a height of about 200~km above the solar surface, in the lower photosphere, where the physics is expected to be dominated by the fluid and not the magnetic field (e.g., Priest~1982).  Fletcher \& Hudson~(2008) argue that large-scale Alfv\'{e}n wave pulses transport energy and magnetic field changes from the flare site rapidly through the corona to the lower atmosphere.  Related theoretical work (Hudson~2000, Hudson et al.~2008) predicts that the flaring magnetic fields undergo an implosion or inward contraction and become more horizontal as a result of flares.

In this paper, we extend the work of SH05 from a sample of 15 X-class flares that occurred over a period of two years to a total of 77 flares - 38 X-class flares and 39 M-class flares down to M5.0 - that occurred over a period of six years.  As in SH05, we characterize the abrupt, longitudinal magnetic field changes associated with the flares at representative points in each active region.  We report on the strength and duration of the field changes and the time of the field changes with respect to the GOES X-ray signatures of the flares.  We also test for correlations between the field changes and background field strength, GOES X-ray flux, and position on the solar disk.  We report separate statistics for subsets of the data set, partitioning the data according to GOES peak X-ray flux (X-class/M-class), strength of the field change (greater than/less than 100 G), and position on the solar disk (near-disk-center/near-limb).

Whereas SH05 limited their study to field changes at representative points in each active region, we go one step further and calculate the change in the longitudinal magnetic flux over the entire active region and characterize these changes in a manner similar to the field changes.  The magnetic flux may be a much more meaningful physical quantity for many purposes than the change in the field strength at any one location but more complicated because of the noise inherent in the data.  

SH05 stated the expectation that the flux changes might correlate with GOES peak X-ray flux but did not test this correlation.  We do so here, testing the correlation between both the field and flux changes and the GOES peak X-ray flux.  Any significant correlation would suggest that the energetics of the X-ray emission and the energetics of the field/flux changes are related.

SH05 reported no significant correlation between field change and position on the solar disk.  We check this result with our larger data set.  Because we measure the longitudinal component of the photospheric field, whose tilt angle with respect to the surface varies as a simple function of position on the disk, any correlation between the field/flux changes and position on the disk tells us which component of the magnetic field tends to change most during the flare, assuming that the field changes in direction and not in magnitude.

Finally, we estimate the Lorentz forces associated with the field/flux changes using A. N. McClymont's incisive method (Anwar et al. 1993, Hudson et al. 2008).  Based on estimates of the total magnetic flux change during a typical X-class flare, Hudson et al. (2008) estimated that changes in the photospheric field due to such flares might be energetically important for seismic waves.  Here we calculate forces corresponding to our measured field changes to see how energetically important changes in the photospheric field are in general.

The paper is organized as follows.  The data are described in Section~\ref{data} and the analysis techniques in Section~\ref{dataanalysis}.  The field and flux changes of the 77 flares are characterized in Sections~\ref{fieldchanges} and \ref{fluxchanges}.  Correlation of field and flux changes with GOES peak X-ray flux is investigated in Section~\ref{fieldfluxint}, and the dependence of field and flux changes on position on the solar disk is explored in Section~\ref{diskposition}.  Estimates of the changes in Lorentz force during the flares are derived in Section~\ref{forces}.  We discuss the implications of our work in Section~\ref{discussion} and conclude in Section~\ref{conclusion}.

\section{The data}
\label{data}

\begin{figure*}[ht]
\begin{center}
\resizebox{\hsize}{!}{\includegraphics*{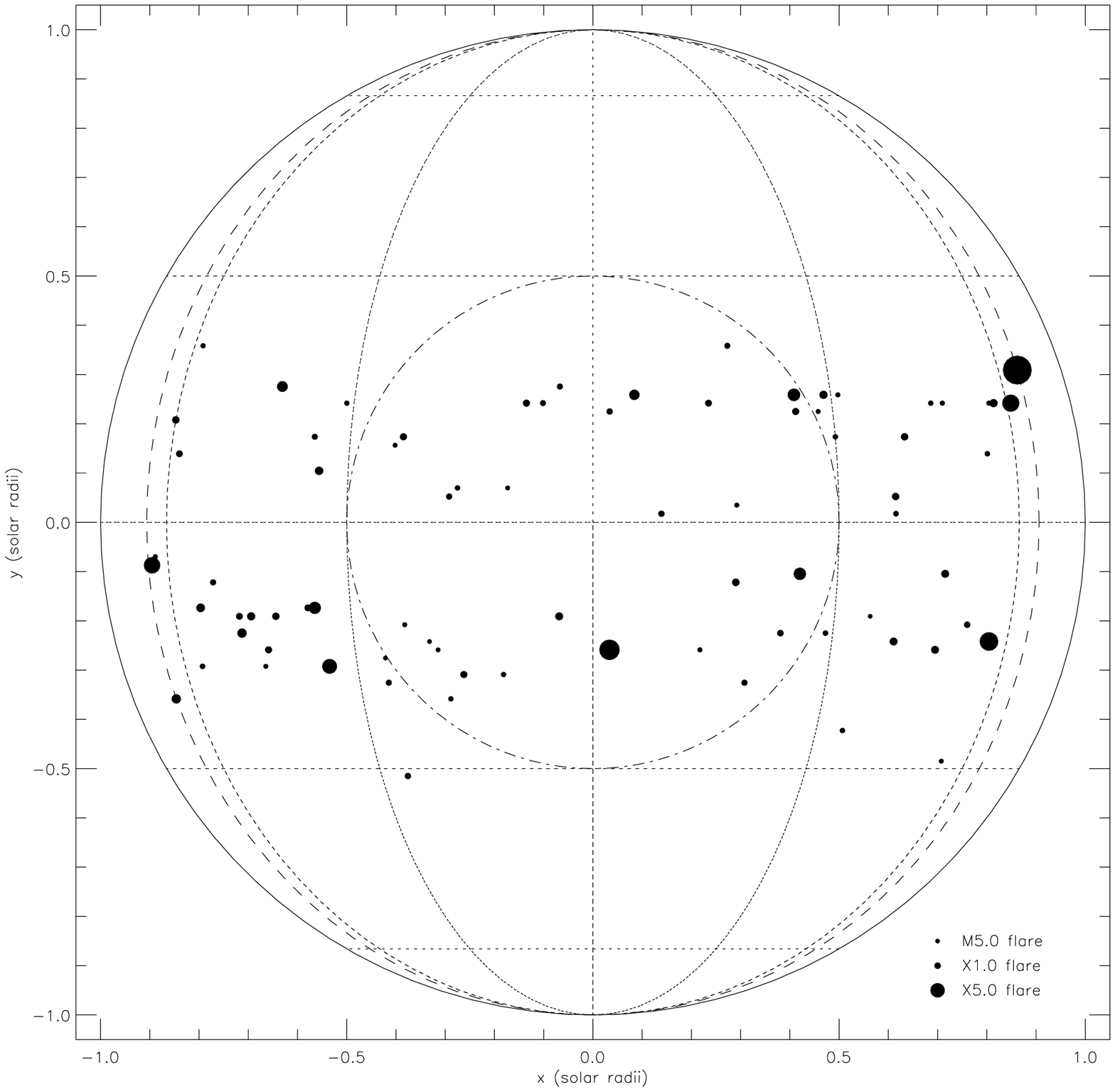}}
\end{center}
\caption{The positions of the 77 flares on the solar disk.  The GOES peak X-ray flux is represented by circle size.  The short-dashed lines mark $0^{\circ}$, $\pm 30^{\circ}$ and $\pm 60^{\circ}$ longitude and latitude.  Only flares located within $65^{\circ}$ of central meridian, within the long-dashed lines, are studied.  The dot-dash circle $r=r_s/2$ separates locations deemed in this study to be near disk-center ($r \le r_s/2$) and those near the limb ($r > r_s/2$), where $r_s$ is the solar disk radius in the image plane.}
\label{diskpos}
\end{figure*}

\begin{table}
\scriptsize
\caption{M-class flares studied in this survey.}
\begin{tabular}{lcccc}
\hline\hline
Date (UT) & {\it GOES} Start Time & {\it GOES} Class & Location & NOAA Number \\
\hline
2001 Jun 22 & 2214 & M6.2 & N14W47 & 9503\\
2001 Jun 23 & 0010 & M5.6 & N09E24 & 9511\\
2001 Sep 5 & 1425 & M6.0 & N15W31 & 9601\\
2001 Sep 9 & 2040 & M9.5 & S31E26 & 9608\\
2001 Sep 16 & 0339 & M5.6 & S29W54 & 9608\\
2001 Oct 22 & 1427 & M6.7 & S21E18 & 9672\\
2001 Oct 23 & 0211 & M6.5 & S18E11 & 9672\\
2001 Nov 7 & 1930 & M5.7 & S17E44 & 9690\\
2001 Nov 8 & 0659 & M9.1 & S19W19 & 9687\\
2001 Nov 28 & 1626 & M6.9 & N04E16 & 9715\\
2001 Nov 29 & 1012 & M5.5 & N04E10 & 9715\\
2001 Dec 26 & 0432 & M7.1 & N08W54 & 9742\\
2002 Jan 9 & 1742 & M9.5 & N13W02 & 9773\\
2002 Mar 14 & 0138 & M5.7 & S12E23 & 9866\\
2002 Jul 11 & 1444 & M5.8 & N21E58 & 10030\\
2002 Jul 17 & 0658 & M8.5 & N21W17 & 10030\\
2002 Jul 26 & 2051 & M8.7 & S19E26 & 10044\\
2002 Aug 16 & 1132 & M5.2 & S14E20 & 10069\\
2002 Aug 20 & 0133 & M5.0 & S11W35 & 10069\\
2002 Oct 5 & 2042 & M5.9 & N14E31 & 10139\\
2002 Nov 18 & 0201 & M7.4 & S17E56 & 10198\\
2002 Dec 20 & 1313 & M6.8 & S25W34 & 10226\\
2003 Oct 26 & 2134 & M7.6 & N01W38 & 10484\\
2003 Oct 27 & 0921 & M5.0 & S16E26 & 10486\\
2003 Nov 20 & 0735 & M9.6 & N01W08 & 10501\\
2003 Nov 20 & 2342 & M5.8 & N02W17 & 10501\\
2004 Jan 17 & 1735 & M5.0 & S15E19 & 10540\\
2004 Jan 20 & 0729 & M6.1 & S15W13 & 10540\\
2004 Jul 13 & 0009 & M6.7 & N14W45 & 10646\\
2004 Jul 13 & 1924 & M6.2 & N14W56 & 10646\\
2004 Jul 20 & 1222 & M8.6 & N10E35 & 10652\\
2004 Jul 22 & 0014 & M9.1 & N03E17 & 10652\\
2004 Jul 25 & 0539 & M7.1 & N10W30 & 10652\\
2004 Aug 14 & 0536 & M7.4 & S13W29 & 10656\\
2004 Oct 10 & 1618 & M5.9 & N13W28 & 10656\\
2005 Jan 15 & 0426 & M8.4 & N14E06 & 10720\\
2005 Jan 15 & 0554 & M8.6 & N16E04 & 10720\\
2006 Dec 6 & 0802 & M6.0 & S04E63 & 10930\\
2007 Jun 4 & 0506 & M8.9 & S07E51 & 10960\\
\hline
\end{tabular}
\label{mtable}
\end{table}

\begin{table}
\scriptsize
\caption{X-class flares studied in this survey.}

\begin{tabular}{lcccc}
\hline\hline
Date (UT) & {\it GOES} Start Time & {\it GOES} Class & Location & NOAA Number \\
\hline
$^*$2001 Apr 2 & 2132 & X20.0 & N18W65 & 9393\\
$^*$2001 Jun 23 & 0402 & X1.2 & N10E23 & 9511\\
$^*$2001 Aug 25 & 1623 & X5.3 & S17E34 & 9591\\
$^*$2001 Oct 19 & 1613 & X1.6 & S17E34 & 9661\\
$^*$2001 Oct 22 & 1744 & X1.2 & S18E16 & 9672\\
$^*$2001 Dec 11 & 0758 & X2.8 & N16E41 & 9733\\
$^*$2002 May 20 & 1521 & X2.1 & S21E65 & 9961\\
$^*$2002 Aug 21 & 0528 & X1.0 & S12W51 & 10069\\
2003 Mar 17 & 1850 & X1.5 & S14W39 & 10314\\
2003 Mar 18 & 1151 & X1.5 & S15W46 & 10314\\
$^*$2003 May 27 & 2256 & X1.3 & S07W17 & 10365\\
$^*$2003 May 28 & 0017 & X3.6 & S06W25 & 10365\\
$^*$2003 Jun 10 & 2319 & X1.3 & N10W40 & 10375\\
$^*$2003 Jun 11 & 2001 & X1.6 & N14W57 & 10375\\
2003 Oct 19 & 1629 & X1.1 & N08E58 & 10484\\
$^*$2003 Oct 26 & 0557 &  X1.2 & S15E43 & 10486\\
2003 Oct 26 & 1721 & X1.2 & N02W38 & 10484\\
$^*$2003 Oct 29 & 2037 & X10.0 & S15W02 & 10486\\
$^*$2003 Nov 2 & 1703 & X8.3 & S14W56 & 10486\\
2004 Feb 26 & 0150 & X1.1 & N14W14 & 10564\\
2004 Jul 15 & 0130 & X1.8 & S10E34 & 10649\\
2004 Jul 15 & 1815 & X1.6 & S11E45 & 10649\\
2004 Jul 16 & 0143 & X1.3 & S11E41 & 10649\\
2004 Jul 16 & 1032 & X1.1 & S10E36 & 10649\\
2004 Jul 16 & 1349 & X3.6 & S10E35 & 10649\\
2004 Aug 13 & 1807 & X1.0 & S13W23 & 10656\\
2004 Oct 30 & 1138 & X1.2 & N13W25 & 10691\\
2005 Jan 1 & 0001 & X1.7 & N06E34 & 10715\\
2005 Jan 15 & 0022 & X1.2 & N14E08 & 10720\\
2005 Jan 15 & 2225 & X2.6 & N15W05 & 10720\\
2005 Jan 17 & 0659 & X3.8 & N15W25 & 10720\\
2005 Jan 20 & 0636 & X7.1 & N14W61 & 10720\\
2005 Jul 30 & 0617 & X1.3 & N12E60 & 10792\\
2005 Sep 10 & 1634 & X1.1 & S11E47 & 10808\\
2005 Sep 10 & 2130 & X2.1 & S13E47 & 10808\\
2005 Sep 13 & 1919 & X1.5 & S05E15 & 10808\\
2006 Dec 6 & 1829 & X6.5 & S05E64 & 10930\\
2006 Dec 14 & 2107 & X1.5 & S06W46 & 10930\\
\hline
$^*$Flares studied by SH05.\label{xtable}
\end{tabular}
\end{table}

Changes in the magnetic field during a solar flare occur on a timescale of 10 minutes.  Photospheric line profile changes occur over a few minutes, and non-flaring active-region fields can evolve at a rate of a few gauss per minute (SH05).  Hence we need of order an hour of uninterrupted high-sensitivity, high-cadence data to distinguish field changes associated with a flare from other changes in the field.

Full-disk images of the relative Doppler shift of the Ni~{\sc I} line at 676.8~nm are available from each of the six GONG telescopes at a cadence of one minute, weather permitting.  GONG's six stations together provide round-the-clock coverage with approximately an 87\% duty cycle.  The spatial sampling of the GONG images is $2.\!\!^{\prime\prime}5$~pixel$^{-1}$ and the instrumental sensitivity is about 3~G~pixel$^{-1}$.  The GONG magnetograms therefore provide the magnetic sensitivity, high cadence, spatial resolution, and spatial and temporal coverage required for the study of magnetic field changes during flares.  The magnetograms are derived from the difference between one-second interleaved observations in right- and left-polarized light and their pixel values are given in meters per second.  These are scaled to Gauss using the factor 0.352~G~m$^{-1}$~s$^{-1}$.

We have analyzed 83 sets of GONG magnetograms for 38 X-class flares, including the 15 examined by SH05, and 39 M-class flares.  Six flares were observed by two sites simultaneously.  Between April and October 2001 the instruments at the six GONG sites were upgraded to the current spatial scale of $2.\!\!^{\prime\prime}5$~pixel$^{-1}$ and full-time magnetic measurements began.  Of the hundreds of M- and X-class flares between then and the last major flares of Solar Cycle 23, we limited our attention to the most energetic flares with the best data coverage.  In particular, we eliminated all flares weaker than M5.0 and all flares with an apparent central meridian longitude difference greater than 65$^{\circ}$ ($\mu\approx 0.42$).  We further limited our attention to those flares for which GONG magnetograms are available of order one hour before and after the flare.  The flares studied are identified in Tables~\ref{mtable} (M-class) and \ref{xtable} (X-class).

Figure~\ref{diskpos} shows the locations of the flares on the solar disk.  These locations derive from the GOES X-ray flare catalog and they are also listed in Tables~\ref{mtable} and \ref{xtable}.  They are fairly evenly spread across the active belt of the Sun, between about $\pm 30^{\circ}$ of latitude.

\section{Data Analysis}
\label{dataanalysis}

\begin{figure*}[ht]
\begin{center}
\resizebox{0.75\hsize}{!}{\includegraphics*{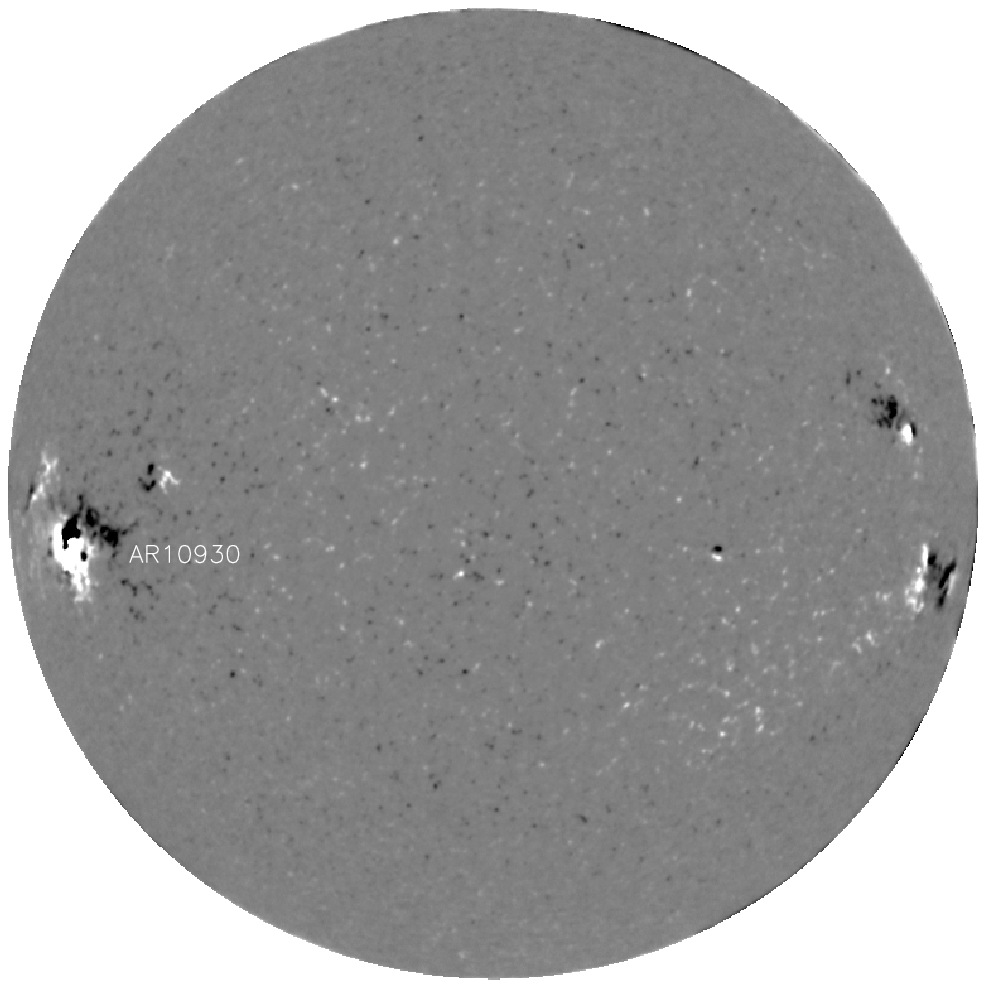}}
\end{center}
\caption{The average of the 10 1-minute GONG longitudinal full-disk magnetograms immediately before the X6.5 flare observed on December 6th 2006 at 1829 UT in Active Region 10930.}
\label{fulldisk}
\end{figure*}

\begin{figure*}[ht]
\begin{center}
\resizebox{0.49\hsize}{!}{\includegraphics*{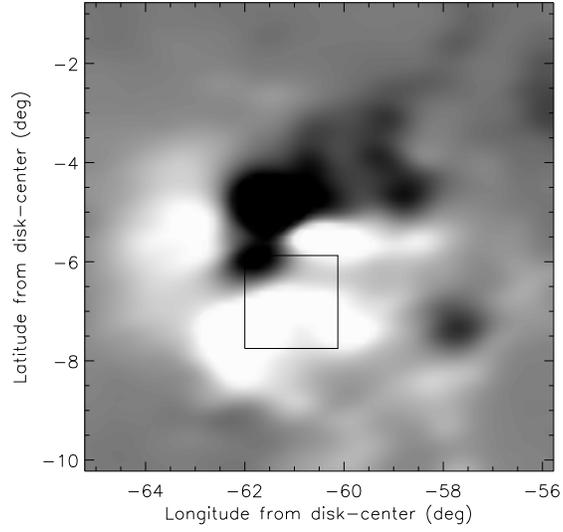}}
\end{center}
\caption{Remapped image of the longitudinal magnetic field of AR10930 based on 10 GONG images immediately preceding the X6.5 flare observed on December 6th 2006 at 1829 UT.  The black square corresponds to the pixels featured in the mosaic plot in Figure~\ref{mosaic}.}
\label{remap}
\end{figure*}

We remapped the active region associated with each flare as in SH05.  We remapped each full-disk image to local heliographic coordinates on the plane tangent to a point near the center of the flaring active region using fourth-order spline interpolation.  The remapped images are $256\times 256$ pixels in size and represent and field of view of $32^{\circ}\times 32^{\circ}$ in heliographic coordinates.  We registered every remapped magnetogram to a reference image formed from the average of the 10 remapped magnetograms immediately preceding the flare.  A full-disk image for the 2006 December 6 X6.5 flare is shown in Figure~\ref{fulldisk}.  This example is close to the 65$^{\circ}$ limit that we impose on the data.  The registration reference image for the 2006 December 6 X6.5 flare is shown in Figure~\ref{remap}.  To first order, the registration corrects for any drift of the active region with respect to the heliographic center of the frame and for any residual error in the orientation of solar north in the images.  These shifts are executed as in SH05 by minimizing the difference between the square root of the absolute value of each frame and that of the reference image.

From each time series of remapped images, we constructed a time series of the field strength of each pixel for up to four hours, two hours before and after the start of the flare.  As in SH05, we fit the function

\begin{equation}
B_l(t)= B_{lin} (t) + B_{step} (t) ,
\label{atancurve}
\end{equation}

\noindent where $B_{lin}(t) = a+bt $ and

\begin{equation}
B_{step}(t) = c\left\{ 1+\frac{2}{\pi} \tan^{-1} \left[ n(t-t_0)\right]\right\} \label{bstep} ,
\end{equation}

\noindent to each time series.  Here $t$ represents time, $a$ and $b$ model the background field evolution, $c$ represents the half-amplitude of the field change, $t_0$ represents the midpoint of the field change, and $n$ is the inverse of the timescale over which the field change occurs.  The rate of field change is

\begin{equation}
\left.\frac{dB_{step}}{dt}\right|_{t=t_0} = \frac{2 c  n}{\pi} ,
\label{stepderiv}
\end{equation}

\noindent where the field change and the duration of the change are given by $dB_l=2c$ and $dt=\pi /n$, respectively.  Like SH05, we simply characterize the field changes in this way without providing a physical model.

\begin{figure*}[ht]
\begin{center}
\resizebox{0.75\hsize}{!}{\includegraphics*{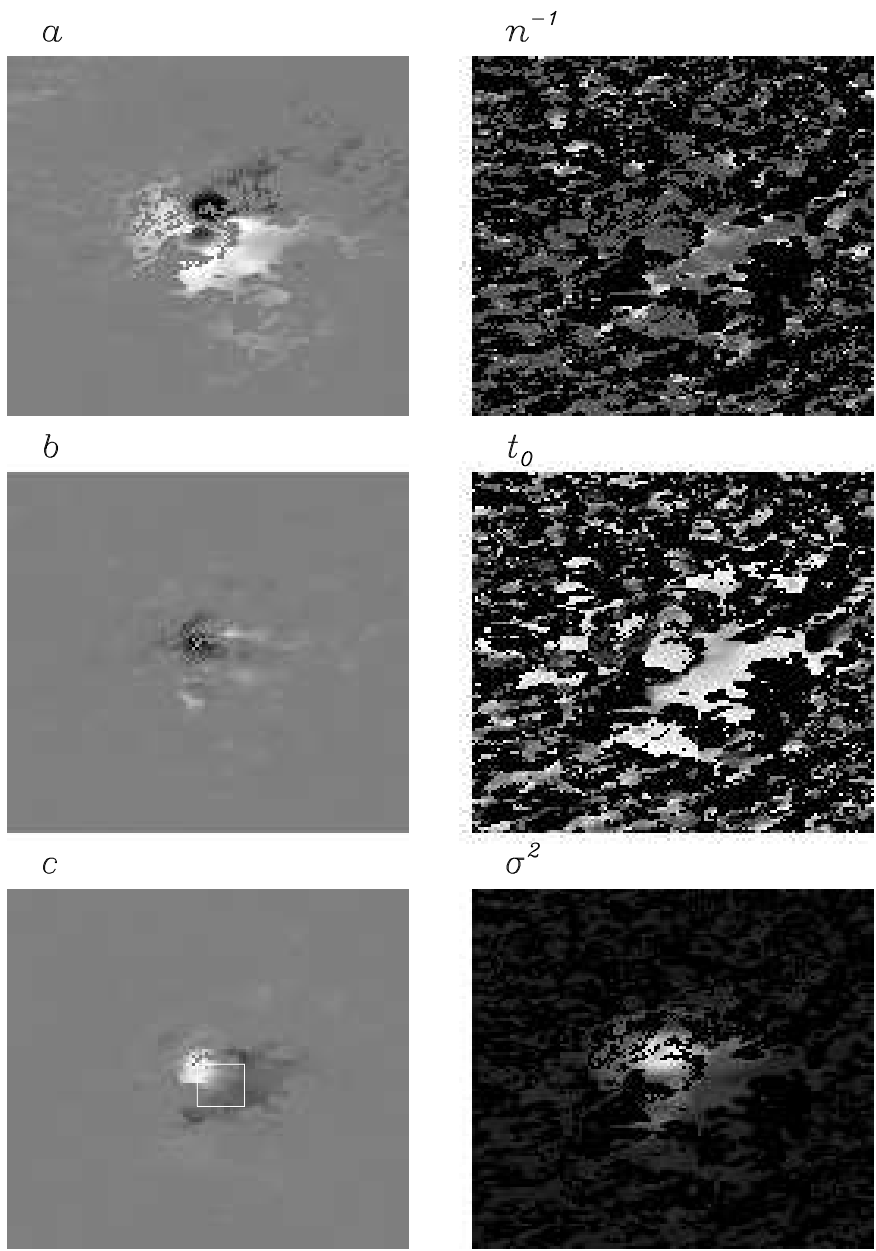}}
\end{center}
\caption{Shown are parameter maps, cropped to $128\times 128$ pixels, of the fit parameters $a$, $b$ and $c$ (left column, top to bottom) and $n^{-1}$ and $t_0$ (right column, top and middle) for the X6.5 flare observed on December 6th 2006.  The scatter of the data with respect to the fit, $\sigma^2$, is shown in the bottom right panel.  Excluded data are represented by grey in the left column and black in the right column.  The maps saturate background fields $a$ and abrupt field-changes $2c$ at $\pm 500$~G, linear background field evolution $b$ at $2.5$~G/min and field change durations $\pi n^{-1}$ at 20 minutes.  Field change times $t_0$ range from 100 minutes to 140 minutes and $\sigma^2$ from about 0.17 to 1.89.  The square in the map for parameter $c$ corresponds to the pixels featured in the mosaic plot in Figure~\ref{mosaic}.}
\label{bigarray}
\end{figure*}

From the fits of the function $B_l(t)$ in Equation~\ref{atancurve} to the time series for the pixels we follow SH05 in creating spatial maps for each parameter in Equation~\ref{atancurve}.  An example set of parameter maps for the 2006 December 6 flare is shown in Figure~\ref{bigarray}.  Not all pixels are ultimately included in these maps.  To avoid spurious fits of Equation~(\ref{atancurve}) to flare-emission transients or to noise spikes we include in these maps only those pixels exhibiting reasonably-sized field changes ($|2c|<500$~G) with steps of reasonably short duration ($\pi n^{-1}\le 40$~minutes) and with the time of the step occurring within 20 minutes of the GOES flare start time.  In the example in Figure~\ref{bigarray}, the $a$ and $c$ maps are almost inverses of each other showing in this particular case that most of the field changes reduce the field strength (see also SH05's Figure~2).  The neutral lines of the two maps do not coincide, however, as there is a small region of increasing longitudinal field at the southern tip of the region of negative polarity.  The field changes vary widely in duration and some variation in start times is evident in the $n^{-1}$ and $t_0$ maps, respectively.  Some field changes that occur early in the largest region of positive polarity close to the neutral line appear to propagate south west across this region.  This phenomenon is similar to that observed in the 2001 December 11 flare by SH05.  The $\sigma^2$ map shows the scatter in the data with respect to the fit of Equation~(\ref{atancurve}) to the data.  The scatter (the noise) is greatest where the field is strongest and the field gradient is steepest.  Therefore the $\sigma^2$ map tends to resemble the absolute value of the $a$ map.  This is the case in our example in Figure~\ref{bigarray} except that the strong positive region in the
south-west of the active region does not appear strongly in the $\sigma^2$ map because the noise level
in the South-West is unusually low.

The GONG instrumentation is identical in design
across the network so that images taken simultaneously by two different telescopes should be nearly identical. We compared pairs of parameter maps for each of the six flares observed by two sites simultaneously.  While instrumental differences and differences in seeing conditions inevitably prevent perfect matches between the image pairs, the close resemblance between each pair provides a foundation for confidence in our results.  SH05 already verified that the analysis gave very similar results when applied to GONG and MDI data for the 2003 October 29 flare.

\begin{figure*}[ht]
\begin{center}
\resizebox{0.99\hsize}{!}{\includegraphics*{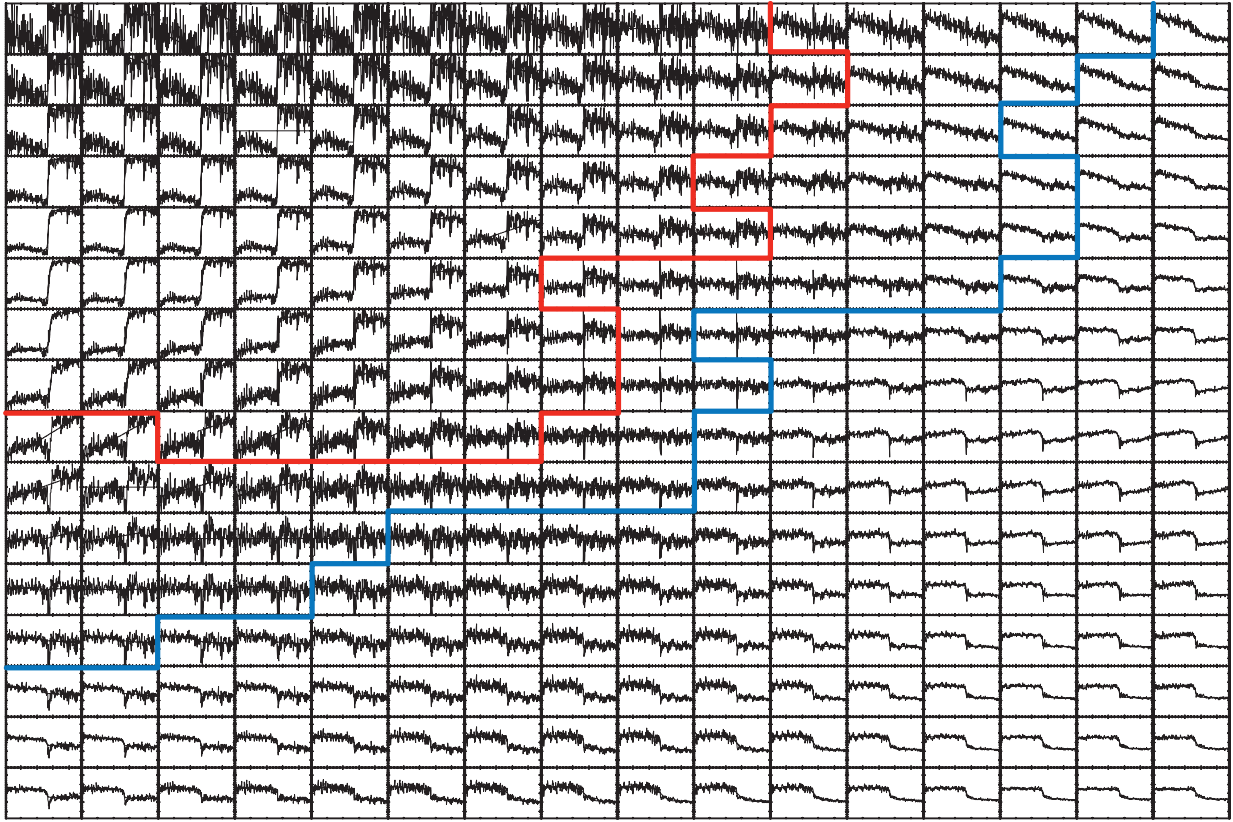}}
\end{center}
\caption{A mosaic of plots of field strength against time for a 16 x 16-pixel subset of the flaring active region 10930 on December 6th 2006 at 1829 UT.  This subset of pixels is indicated by the white solid-line square in the $c$-parameter plot in Figure~\ref{bigarray}.  Each plot in the mosaic corresponds to a single pixel.  The horizontal axis is time and spans 4 hours.  The vertical axis is field strength with its mean value subtracted and spans 500 G.  The red and blue lines mark the boundaries of contiguous regions of positive and negative field changes.  The swath of pixels between the red and blue lines have no significant stepwise field change.}
\label{mosaic}
\end{figure*}

The remapped images for a given flare are stacked to form a space-time data cube.  Figure~\ref{mosaic} shows the evolution of a $16\times 16$ subset of pixels in the form of a $16\times 16$ mosaic of plots of field strength against time arranged to reflect the spatial distribution of the pixels.  In each individual plot of the mosaic, the horizontal axis is time, spanning the four-hour duration of the time series centered at the GOES start time of the flare, and the vertical axis is field intensity.  The region chosen for this particular mosaic straddles the neutral line of the field-change map (parameter $c$) in Figure~\ref{bigarray}.  In the mosaic the neutral line appears as a swath of plots without well-defined stepwise changes, extending from the south east corner to the north west corner of the mosaic.  This swath separates a contiguous region of positive field changes in the North-East, whose boundary is marked with a red line, and a contiguous region of negative changes in the South-West, whose boundary is marked with a blue line.  The positive changes above the left part of the
red line are significantly stronger than anything
reported in SH05. There is also a large, contiguous group
of beautiful, low-noise field changes in the bottom-right of the
mosaic.  Some plots show spikes because of flare-induced line profile changes; the line goes into emission rather than absorption, resulting in unphysical measurements (Edelman et al.~2004).  Examples of this phenomenon include the pixels at the bottom left of the mosaic.  The noise is almost all seeing-related and its strength is sensitive to local intensity gradients and magnetic field gradients.

\clearpage

\clearpage

\clearpage

\clearpage

\section{Field changes}
\label{fieldchanges}

\begin{table}
\scriptsize
\caption{Maximum, minimum and median absolute field changes.}
\begin{tabular}{lccc}
\hline\hline
 & minimum change (G) & maximum change (G) & median change (G) \\
\hline
All flares & 11 & 455 & 69 \\
X-class flares & 11 & 455 & 82 \\
M-class flares & 13 & 236 & 54 \\
Disk-center flares & 13 & 281 & 54 \\
Limb flares & 11 & 455 & 85 \\
Disk-center X-class flares & 16 & 281 & 71 \\
Limb X-class flares & 11 & 455 & 97 \\
Disk-center M-class flares & 13 & 149 & 44 \\
Limb M-class flares & 13 & 266 & 69 \\
\hline
\end{tabular}
\label{minmaxtable}
\end{table}

\begin{table}
\scriptsize
\caption{Field changes: selected statistics, correlations and confidence levels.}
\begin{tabular}{lcccc}
\hline\hline
 & No. fields & No. fields & Pearson c.c. & Probability  \\
& increasing & decreasing & $r_0$ between & $Prob_N (|r|\ge |r_0|)$ \\
&  &  & $B_l$ and $dB_l$ & for $N$ measurements \\
\hline
All field changes & 65 & 94 & -0.035 & 0.66 \\
X-class field changes & 34 & 56 & -0.04 & 0.71 \\
M-class field changes & 31 & 38 & 0.01 & 0.94 \\
Weak field changes & 37 & 71 & -0.43 & $2.4\times 10^{-6}$ \\
Strong field changes & 28 & 23 & 0.20 & 0.17 \\
Disk-center field changes & 23 & 42 & -0.20 & 0.11 \\
Limb field changes & 42 & 52 & 0.037 & 0.72 \\
Weak disk-center field changes & 17 & 36 & -0.51 & $6.9\times 10^{-5}$ \\
Strong disk-center field changes & 6 & 6 & 0.37 & 0.34 \\
Weak limb field changes & 20 & 35 & -0.37 & $5.1\times 10^{-3}$ \\
Strong limb field changes & 22 & 15 & 0.27 & 0.24\\
\hline
\end{tabular}
\label{fieldchangetable}
\end{table}

\begin{figure*}[ht]
\begin{center}
\resizebox{0.65\hsize}{!}{\includegraphics*{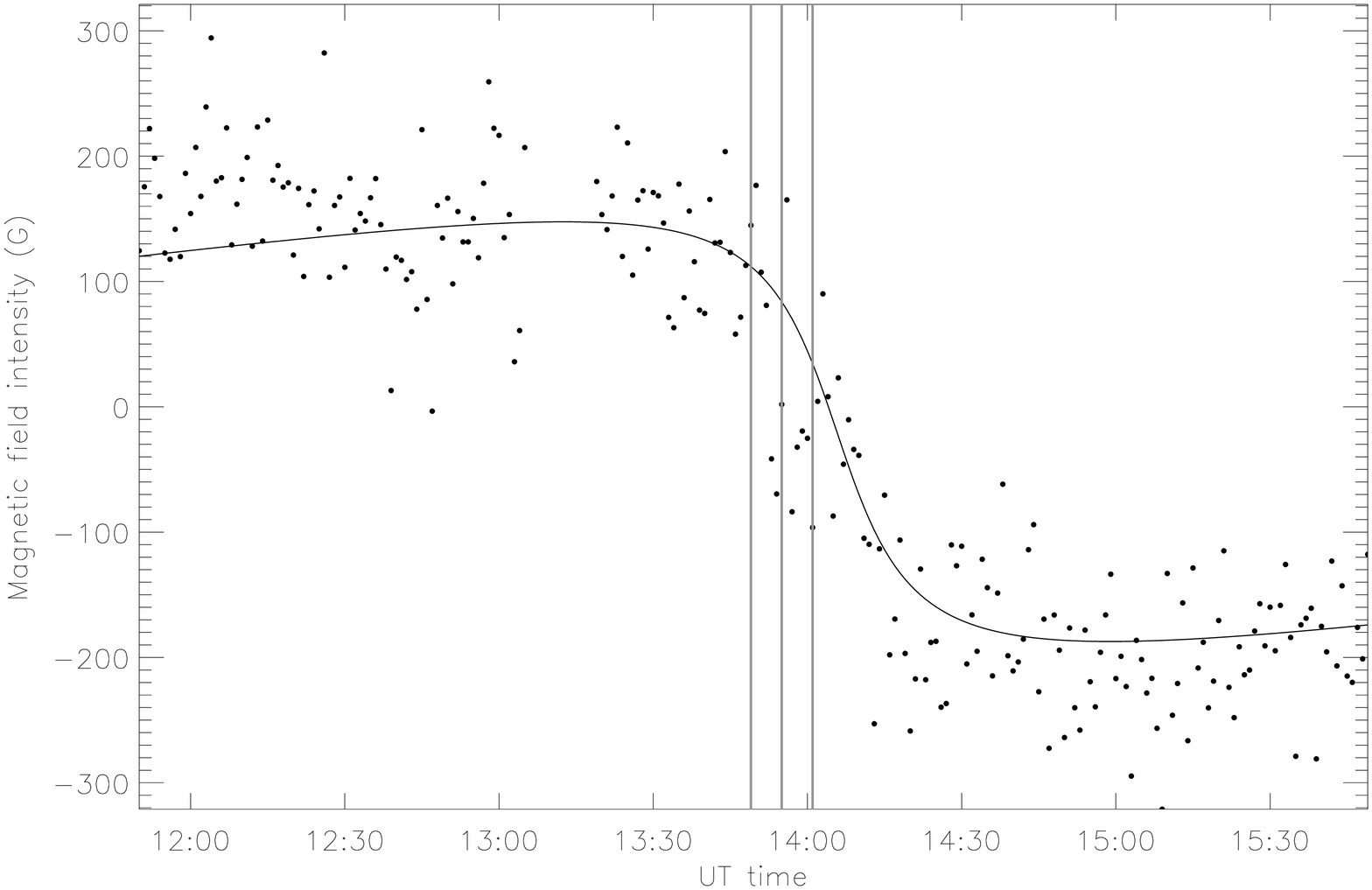}}
\resizebox{0.65\hsize}{!}{\includegraphics*{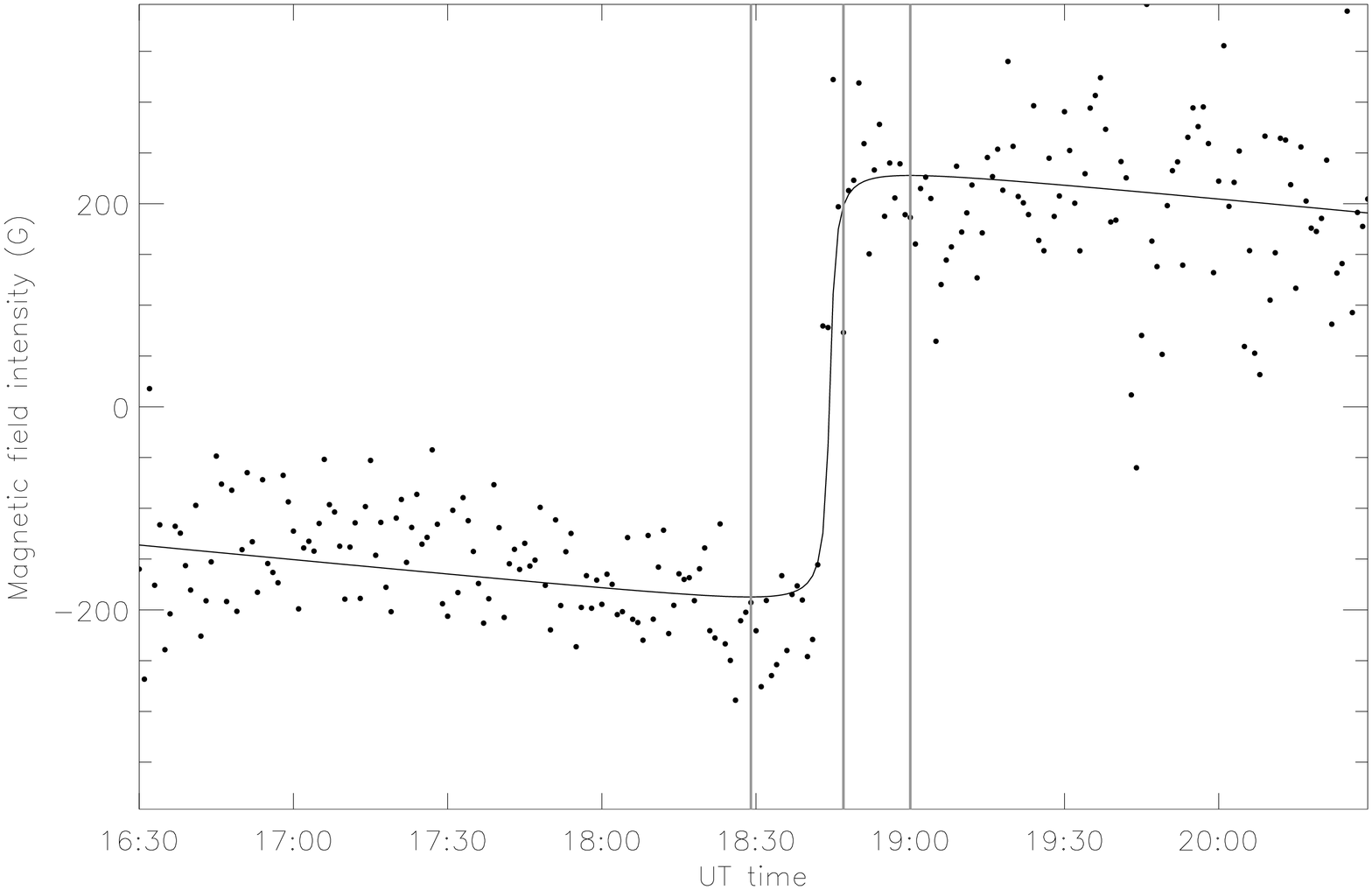}}
\end{center}
\caption{Observational data points and fitted curves for the two strongest field changes in our sample recorded during the 2004 July 16 X3.6 flare (top) and the 2006 December 6 X6.5 flare (bottom).  The mean value of magnetic field intensity has been subtracted from the data and the fit in each plot.  The vertical lines mark the GOES X-ray start, peak and end times of the flares.}
\label{strongcols}
\end{figure*}

The foregoing analysis gives us thousands of stepwise field changes to consider.  Field changes were observed in all 77 flares.  In order to characterize the field changes we followed SH05 in selecting a few representative pixels from each of 77 data sets.  While two good data sets were available for each of six of the flares, we used only one data set per flare in this analysis.  SH05 selected representative pixels by examining by eye over 8000 mosaic plots like the one in Figure~\ref{mosaic}.  This approach on its own is not practical for our larger data set.  To simplify this process we selected for review pixels with the following characteristics: (a) the stepwise change was at least $1.4$ times stronger than the pre-flare background noise level; (b) the time series of measurements passed a reduced-$\chi^2$ test; (c) the background field and field change values were not unreasonably large ($|a|\le 1000$~G, $|2c|\le 350$~G in general); and (d) the field change was complete within 40~minutes.  The
criteria (c) were very helpful in general in eliminating
unconvincing field changes.  All pixels passing the tests (a-d) were then examined by eye and representative pixels chosen.  As Figure~\ref{mosaic} shows, the clarity of the field changes varies greatly.  In some pixels the field change is complicated by background noise and by spikes in the data due to an emission feature in the spectral line during the flare.  For each active region, we tried to represent every significant sub-region of changing flux and to choose the pixels with the strongest, clearest permanent changes free of noise and emission artifacts.  As in SH05, the representative `pixels' are actually averages of the four neighboring pixels that best represent the abruptness and significance of the field change.  The resolution of the GONG magnetograms is about $5^{\prime\prime}$, hence an average of four adjacent pixels best represents the true resolution of the data.  We also examined the field
changes stronger than 350~G by eye and found two sites, one near the 2004 July 16 flare and one near the 2006 December 6 flare, where such strong fields changes clearly took place.  Representative pixels from these two extreme cases appear in Figure~\ref{strongcols}.  Cases with $|a| > 1000$~G existed in our data
set but none was convincing enough to include.  As in SH05, regions in which the background field strength is greater than 1000~G are too noisy for the clear detection of permanent field changes.

In all, we selected 159 representative pixels from the 77 data sets, compared to 43 pixels representing 15 flares in SH05.  The smaller pixel/flare ratio in this study is due
to the inclusion of active regions with simpler magnetic
structure. For example, if only a single magnetic
polarity changed significantly during a given flare we
only chose one pixel to represent it (as occurred in four
cases of SH05 - see their Figure~3). SH05's sample
spanned 2001-2003 and featured several very complex active
regions. Many of the active regions in this study were bipolar, and showed bipolar changes, bringing our
pixel/flare ratio closer to 2.

In each case the field change is permanent in so far as it
persists until the end of the time series. In the vast
majority of cases the time series ends two hours after the
GOES start time of the flare. In some cases the
instrument unstowed or stowed within the four-hour time
interval centered on the GOES start time of the flare. In
all of these cases the time series extends at least an
hour after the field change, and the duration of the field change is much less
than an hour. In this sense all field changes are permanent.

\clearpage

\begin{figure*}[ht]
\begin{center}
\resizebox{0.45\hsize}{!}{\includegraphics*{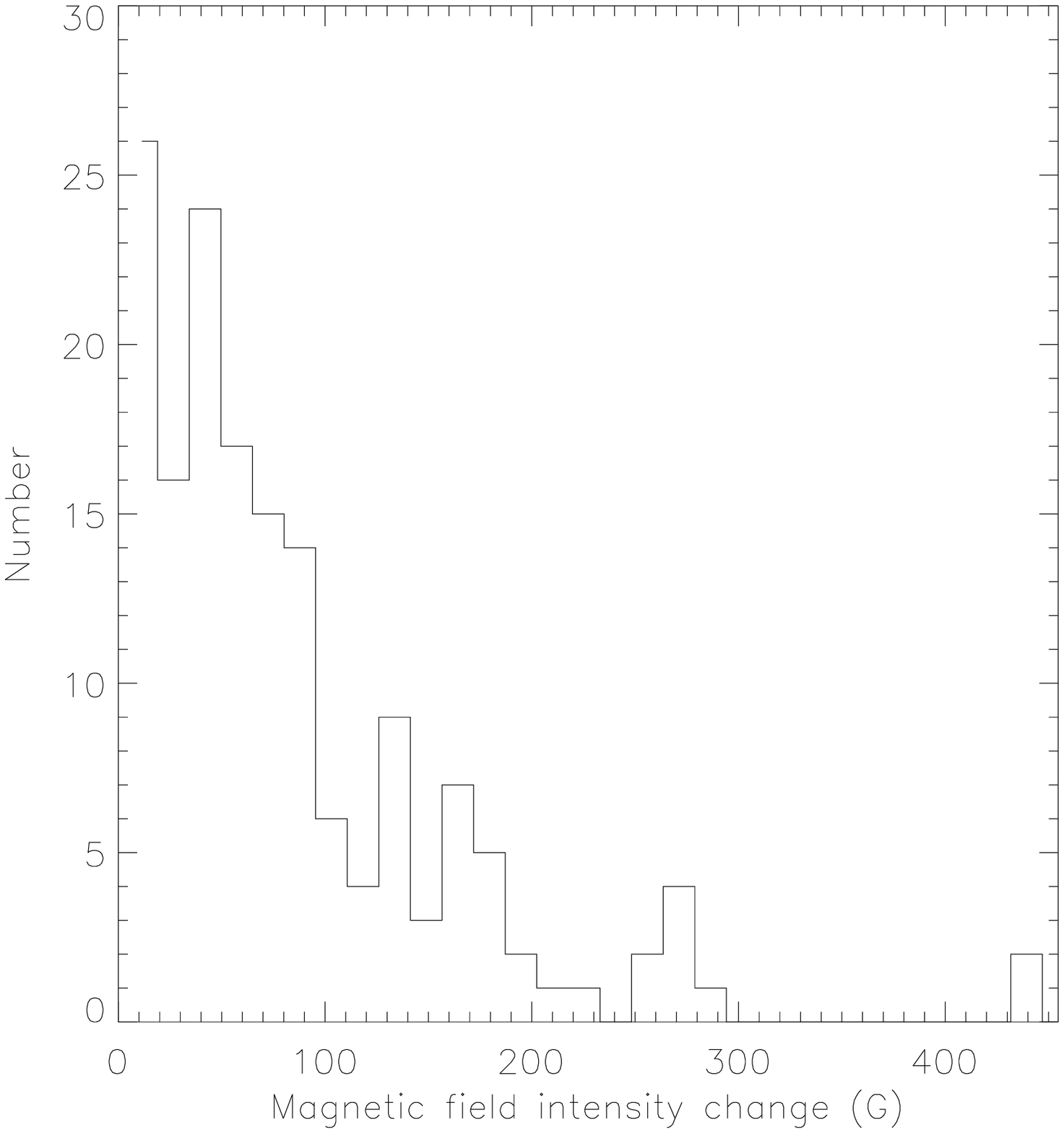}}
\resizebox{0.45\hsize}{!}{\includegraphics*{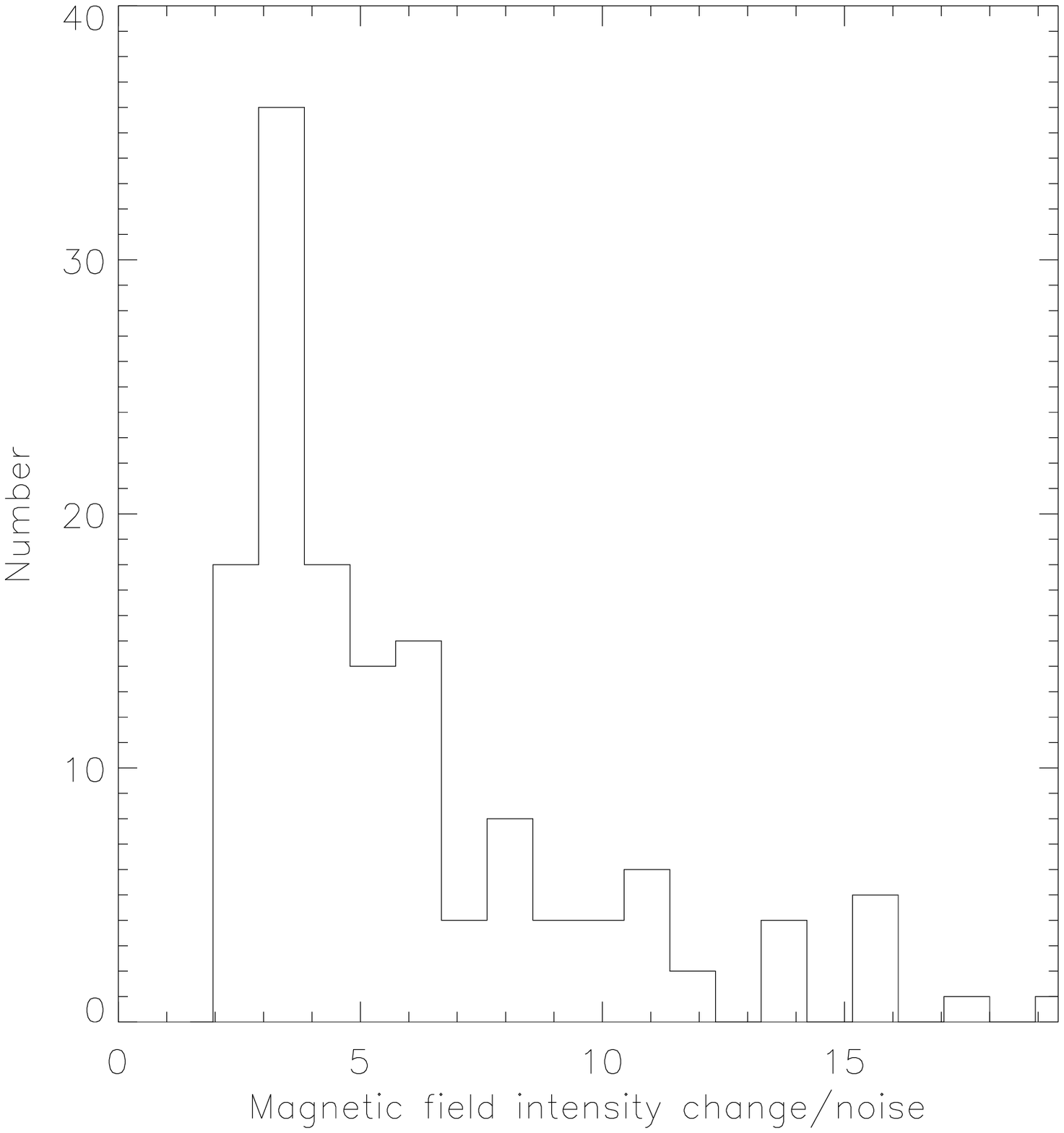}}
\resizebox{0.45\hsize}{!}{\includegraphics*{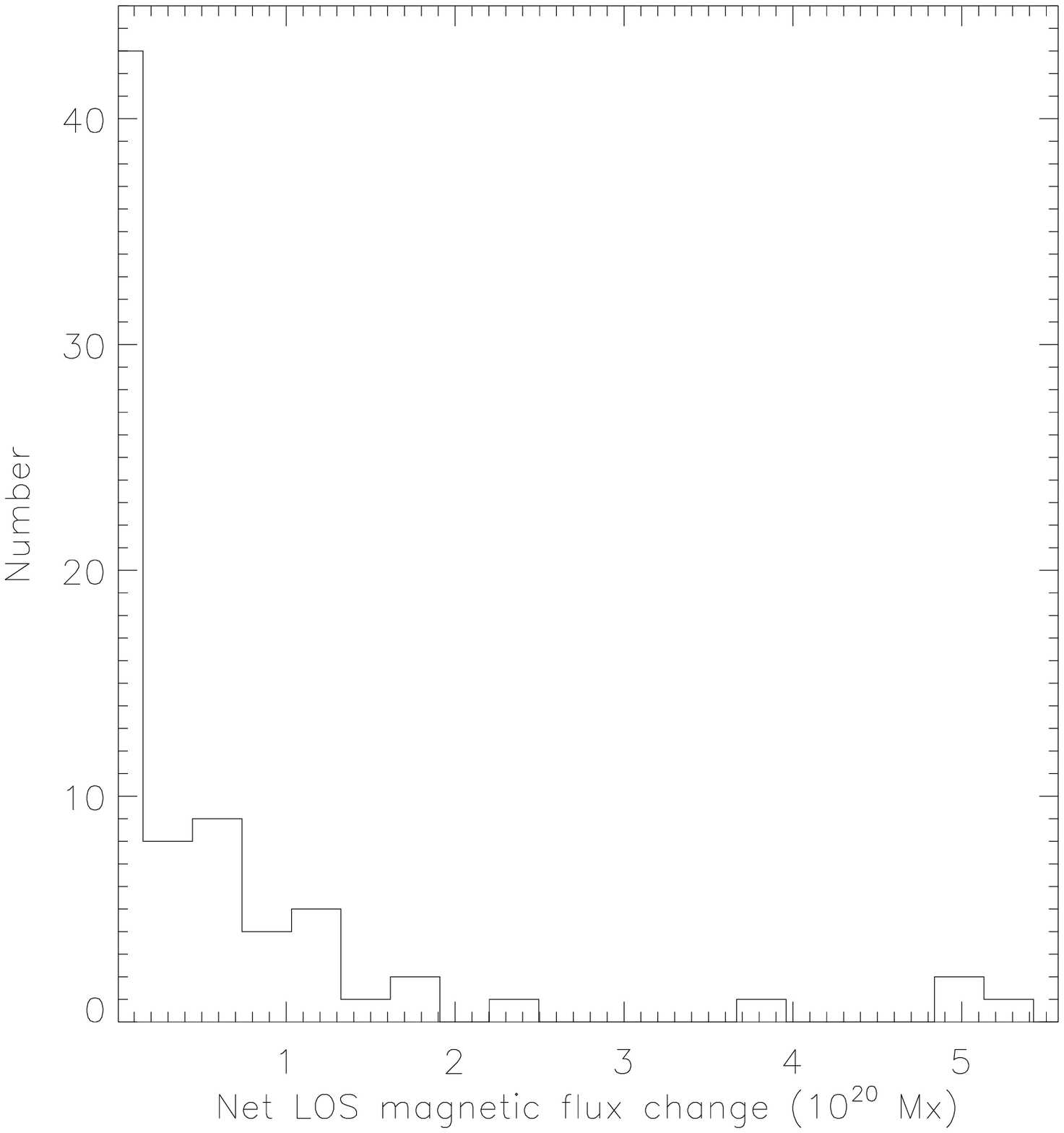}}
\resizebox{0.45\hsize}{!}{\includegraphics*{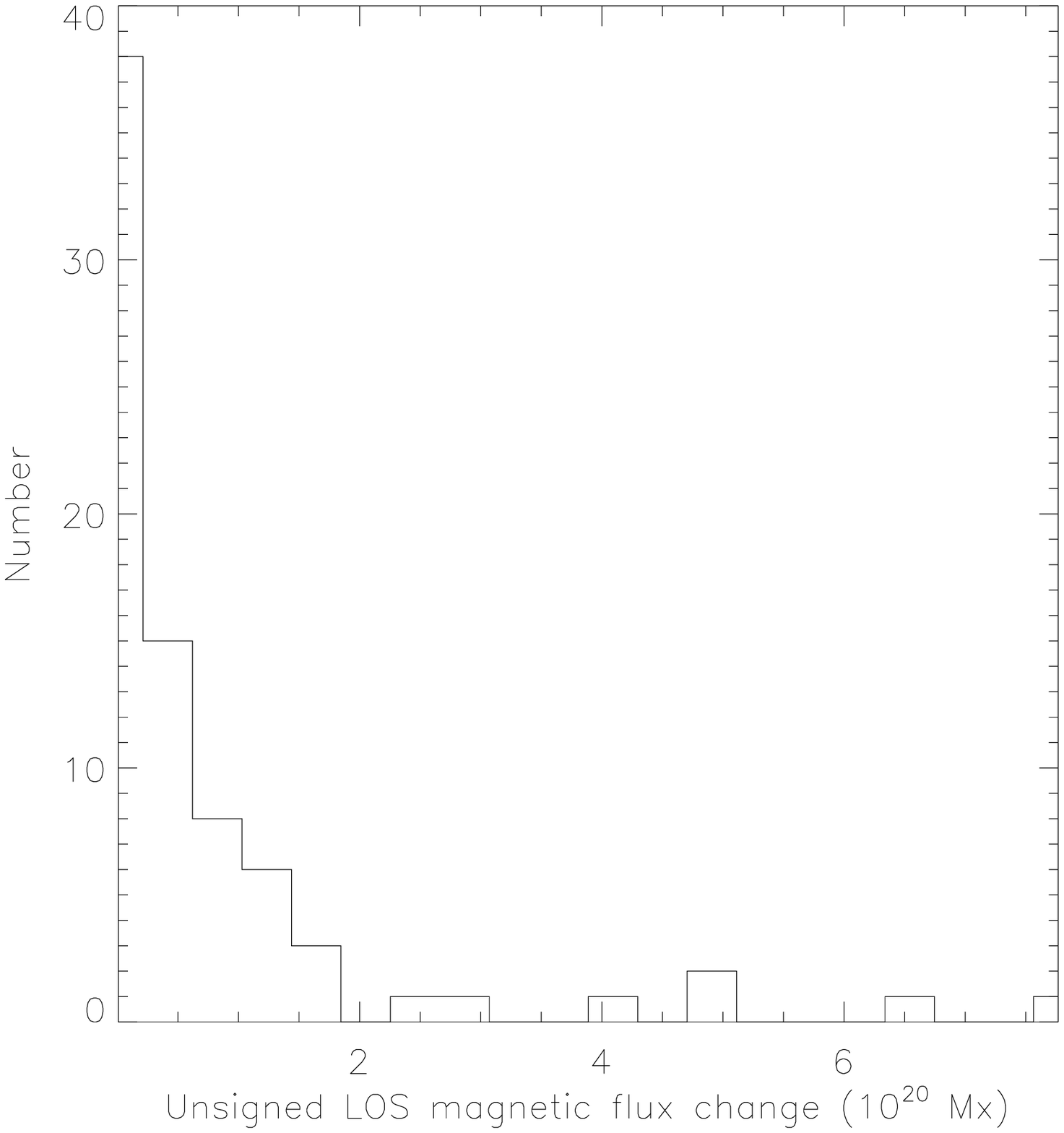}}
\end{center}
\caption{Histograms of field changes (top left), the significance of these field changes (top right), the change in net flux (bottom left) and the change in unsigned flux (bottom right).  The significance is defined as the ratio of the amplitude of the field change to the root-mean-square scatter in the data with respect to the model before the field change occurs.}
\label{fchangehist}
\end{figure*}

The distributions of the field changes are shown at the top of Figure~\ref{fchangehist}.  We will describe the remainder of Figure~\ref{fchangehist} in Section~\ref{fluxchanges}.  The significance of a field change is defined as the ratio of the amplitude of the field change to the root-mean-square scatter of the data with respect to the fit before the field change occurred.  The top right plot of Figure~\ref{fchangehist} shows a histogram of this quantity.  The minimum in the significance plot is due to the fact that we reject field changes with amplitudes less than 1.4 times the noise level, which is around 20~G for strong active fields and around 3~G for quiet fields.  In practice we calculated the noise level
for each pixel under consideration.
The histograms suggest power laws.

\clearpage

\begin{figure*}[ht]
\begin{center}
\resizebox{0.45\hsize}{!}{\includegraphics*{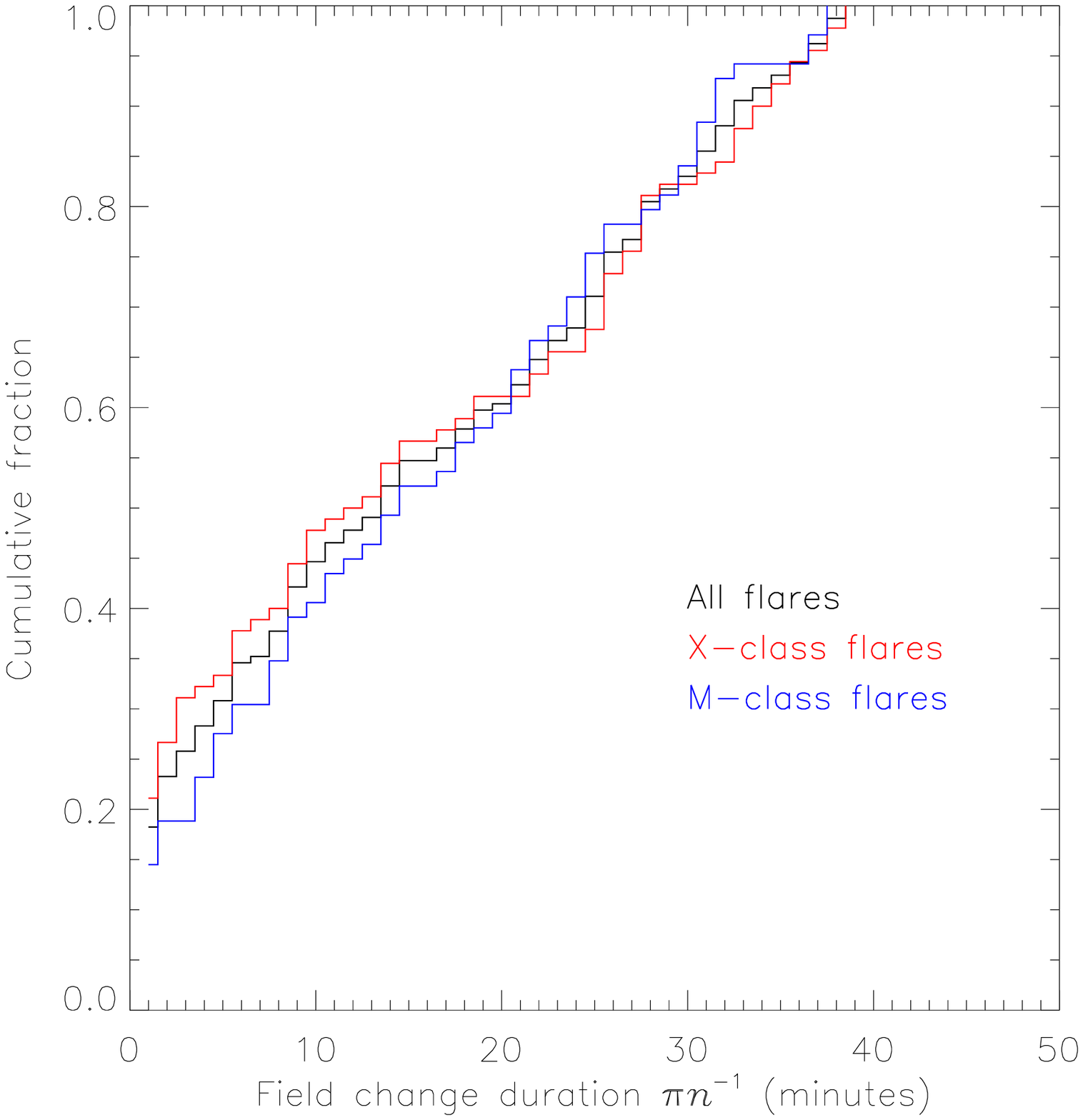}}
\resizebox{0.45\hsize}{!}{\includegraphics*{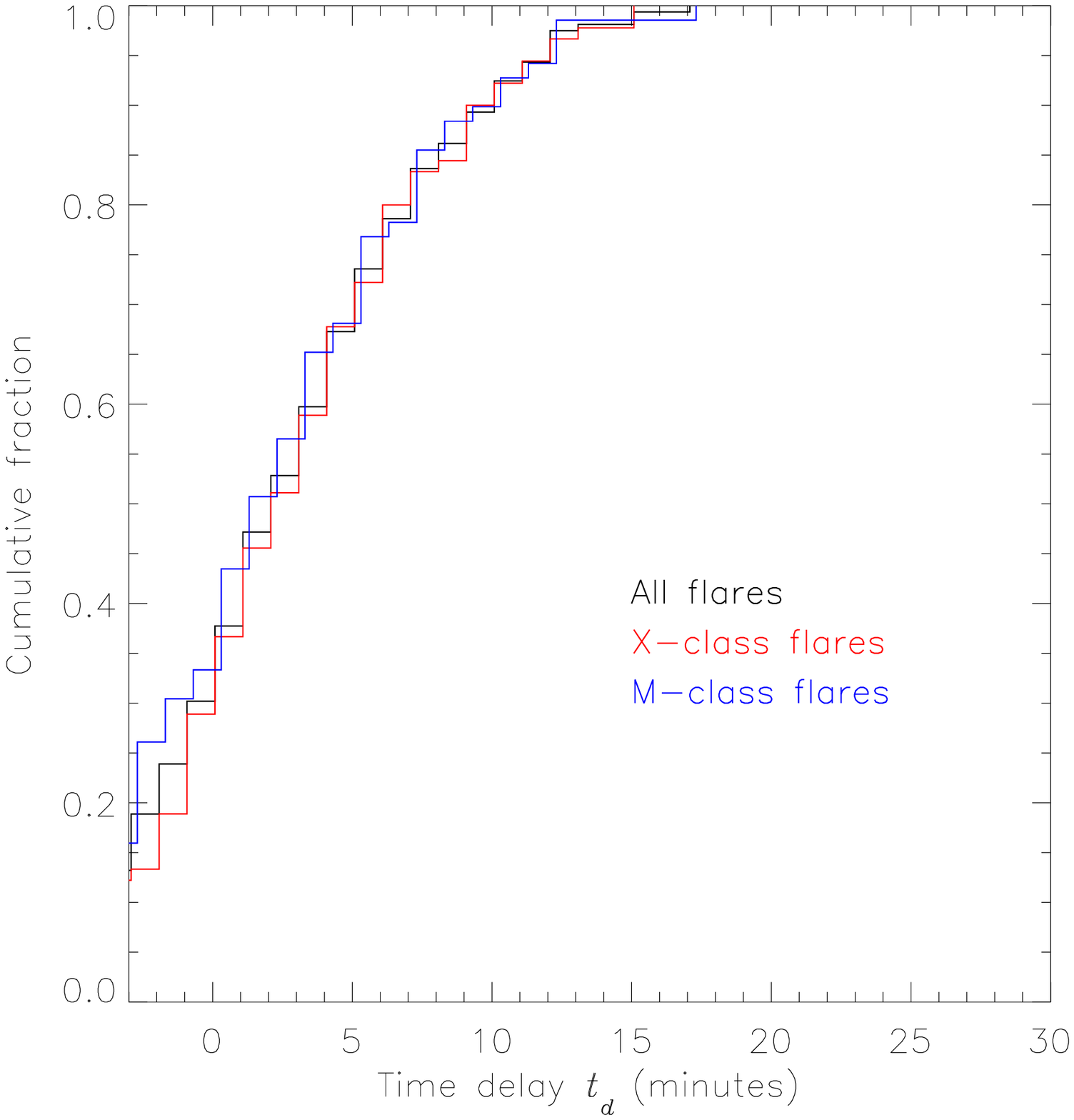}}
\resizebox{0.45\hsize}{!}{\includegraphics*{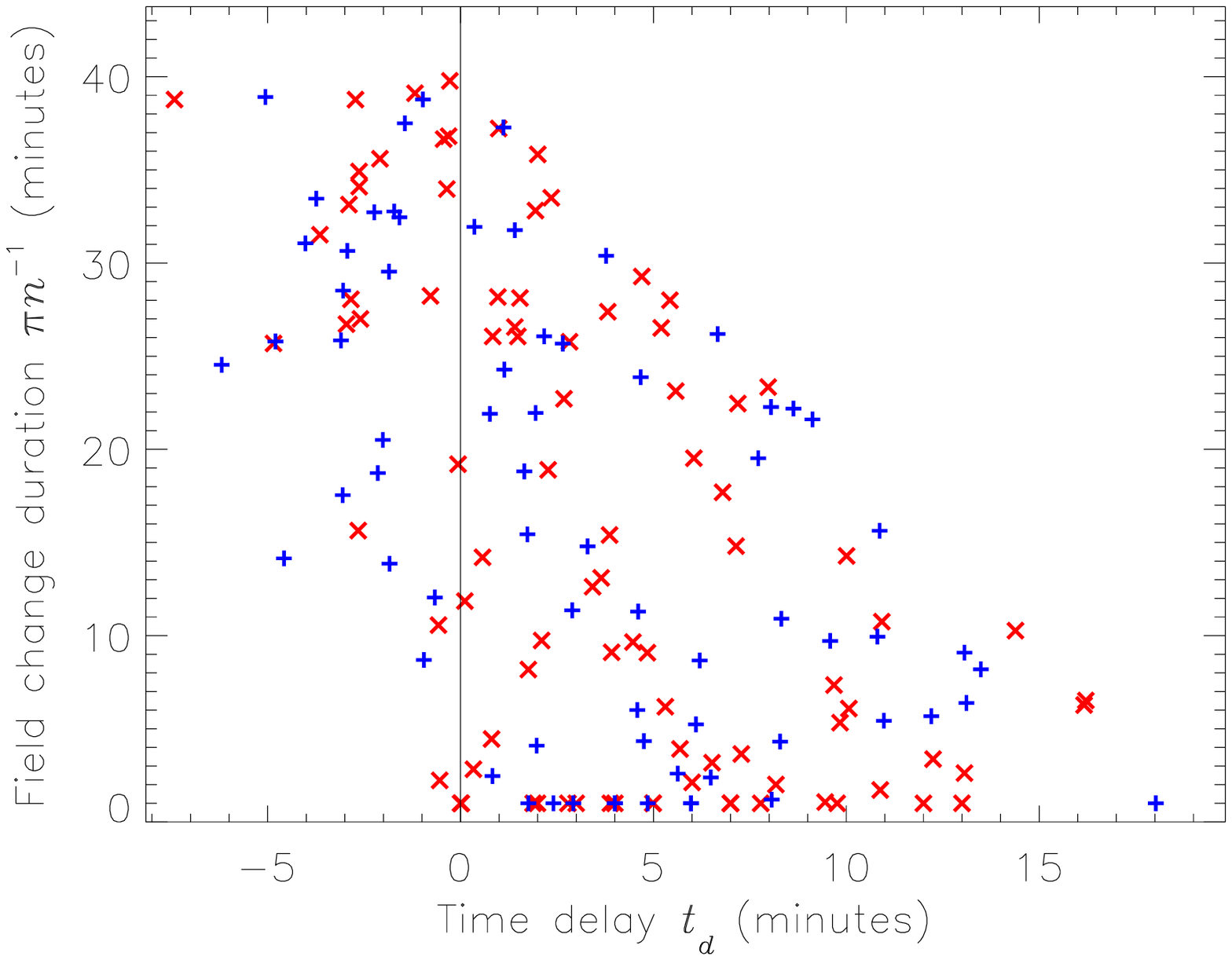}}
\end{center}
\caption{Cumulative histograms of the time periods over which the magnetic field changes occur $\pi n^{-1}$ (top left), cumulative histograms of the time delays between the start times of the GOES X-ray flare signatures and  the magnetic field changes for the X-class and M-class flares separately and all flares combined (top right) and a scatter plot of the time periods over which the field changes occur against time delays (bottom).  In the scatter plot blue plus signs (+) denote M-class flares and red crosses ($\times$) X-class flares.  The vertical line separates cases where the field change start time, derived from the fit of Equation~(\ref{atancurve}) to the data, lags ($t_d>0$) and leads ($t_d<0$) the published GOES X-ray flare start time.}
\label{cumhist}
\end{figure*}

Figure~\ref{cumhist} shows cumulative histograms of the time periods over which the field changes occur, $\pi n^{-1}$, for the X-class and M-class flares separately and for all of the flares combined.  Because the temporal resolution of the data is 1 minute, where $\pi n^{-1} < 1.0$ in the fit to the data, we reset $\pi n^{-1}$ to 1.0 in constructing these histograms.  This occurs in about 20\% of our cases compared to 40\% in SH05.  The difference is due to the inclusion of slower field changes in this work, up to 40 minutes, compared to the upper limit of 20 minutes in SH05.  The time periods over which the field changes occur do not differ significantly between X- and M-class flares.  The median value for the X-class flares is 13 minutes whereas the median value for the M-class flares is 15 minutes.  (The median value for all flares is 14 minutes.)

The second plot of Figure~\ref{cumhist} shows cumulative histograms of the differences between the GOES start times of the flares and the start times of the corresponding magnetic field change for X-class and M-class flares separately and for all of the flares combined.  The start time, $t_{s}$, of the field change is derived from the fit parameters,

\begin{equation}
t_{s} = t_0 - \pi /(2n),
\label{tstart}
\end{equation}

\noindent and the time delay, $t_d$, is the time lag between the GOES X-ray start time of the flare (given in Tables~\ref{mtable} and \ref{xtable}) and the start time of the field change, $t_s$.  In about one third of the cases, the time delays are negative, so it appears that the magnetic field begins to change before the flare occurs.  We emphasize again that Equation~(\ref{atancurve}) does not represent a physical model.  We do not believe that the negative time delays in the second plot in Figure~\ref{cumhist} are meaningful.  The start time of the field change, defined by Equation~(\ref{tstart}), corresponds to the first point of maximum curvature in the step function fit to the data.  The longer the time period over which the field change occurs, the shallower the maximum curvature, and the less certain we can be about the time at which the magnetic field begins to change (see Figure~\ref{strongcols} for two contrasting examples).  Moreover, Equation~(\ref{atancurve}) can represent some measured field changes better than others. For example, if a field change has instantaneous transitions from a constant field to a steady change (straight, sloping graph) to a constant field again, then Equation~(\ref{atancurve}) is doomed to overestimate the field change duration and the estimated start time, $t_s$, is too early. This type of error can occur for both abrupt and gradual field changes but is generally larger for gradual changes.  In other words, the uncertainty in $t_s$ is proportional to the field change duration $\pi n^{-1}$.  Indeed, in our calculations the error in $t_{s}$ is dominated by the error in $n^{-1}$, whose 1-$\sigma$ value is often on the order of a few minutes.  The third plot in Figure~\ref{cumhist} shows the time period over which the field change occurs, $\pi n^{-1}$, against the time delay, $t_d$.  The vertical line separates the field changes that appear to start before the GOES X-ray signature $(t_d < 0$), at least according to the fit of Equation~(\ref{atancurve}) to the data, and the field changes that start after the GOES X-ray signature ($t_d > 0$).  The negative time delays are associated with field changes that occur over long periods of time, more than 10 minutes in most cases.  They occur in proportionally more cases in this study than in SH05 because we include more gradual changes in our data set.  Furthermore, the larger the negative time delay, the longer the duration of the field change.  The negative time delays appear to be an artifact.  We maintain that the magnetic field changes are a consequence of solar flares and not a trigger.

\begin{figure*}[ht]
\begin{center}
\resizebox{0.45\hsize}{!}{\includegraphics*{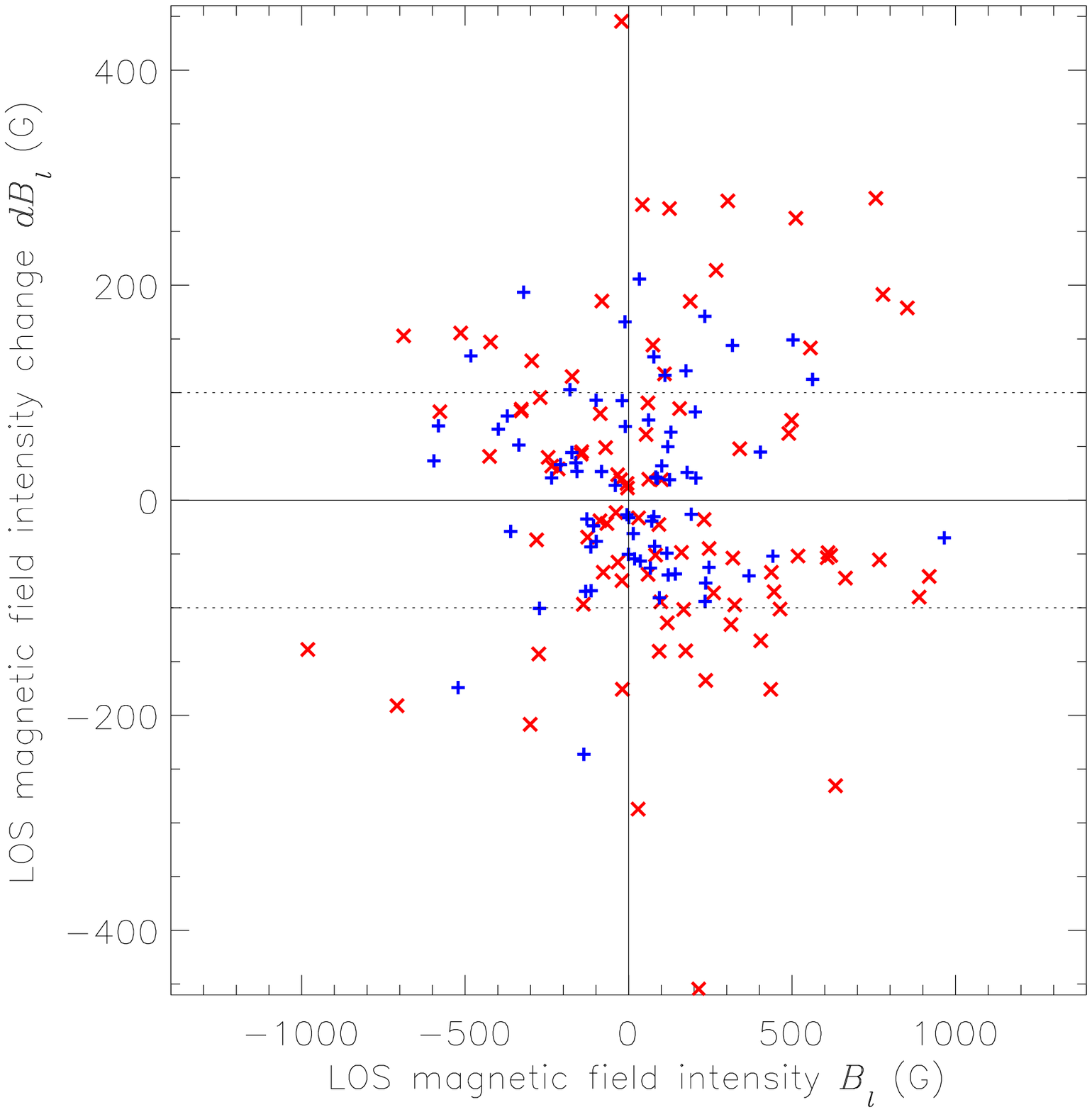}}
\end{center}
\caption{Scatter plot of the longitudinal field changes, $dB_l$, against the background field values, $B_l$.  Blue plus signs (+) denote M-class flares and red crosses ($\times$) X-class flares.  The dotted lines separate weak field changes ($|2c|\le 100$~G) and strong field changes ($|2c|>100$~G).  The former are negatively correlated with background field while the latter are not.}
\label{bvsdb}
\end{figure*}

Figure~\ref{bvsdb} shows a scatter plot of the magnetic field changes, $dB_l$, against background field intensities, $B_l$, for all of the representative pixels in our data set.  The empty, narrow, horizontal stripe represents our detection limit.  Two notable exceptions appear at -455 and +445~G, recorded during the 2004 July 16 X3.6 flare and the 2006 December 6 X6.5 flare, respectively.  Although the parameters of the fits to the time series plots associated with these field changes fall outside our normal criteria for inspection, field changes stronger than 350~G are quite rare, so we were able to examine all of the representative pixels for which $|2c| > 350$~G by eye.  In doing so, we found these two extreme cases, which we present in Figure~\ref{strongcols}.  Apart from these two extreme cases, the distribution of field changes between -300~G and +300~G is consistent with SH05, though we did find field changes as weak as 11~G compared to 28~G in SH05.  

In Table~\ref{minmaxtable}, we summarize the maximum, minimum, and median of the absolute values of the field changes for all of the representative pixels and for different subsets of the data.  The absolute values of all of the field changes range from 11~G to 455~G with a median value of 69~G.  Field changes associated with X-class flares have a median value of 82~G, close to the median value of 90~G reported in SH05, while the M-class flares have a lower median value of 54~G.  SH05 found no correlation between the strength of the field change and position on the solar disk.  Here, however, we do find a split in the data between field changes near the limb and near disk-center.  In this paper we denote by ``near disk-center'' those locations where $r \le r_s/2$ and by ``near the limb'' those where $r > r_s/2$ (see Figure~\ref{diskpos}), where $r_s$ is the solar disk radius in the image plane.  Field changes near the limb have a higher median value, 85~G, than field changes near disk-center, 54~G.  Although there are more X-class flares than M-class flares at the limb, the X-class and limb flare biases are independent.  The median field change for limb X-class flares, 97~G, is significantly higher than both the median disk-center X-class change, 71~G, and the median limb M-class change, 69~G, which are both in turn significantly higher than the median disk-center M-class change, 44~G.

SH05 reported that field changes are twice as likely to decrease the field as increase it.  More precisely, 27 of the 42 observed changes in the longitudinal magnetic field were associated with a decrease in the background field intensity, whereas 15 were associated with an increase, a ratio of 1.8 to 1.  In Table~\ref{fieldchangetable}, we present the statistics for decreasing/increasing fields.  Of the 159 field changes in our study, 94 decreased the field while 65 increased it, a ratio of 1.4 to 1, lower than in SH05.  The asymmetry is apparent in Figure~\ref{bvsdb}; the top left and bottom right quadrants are more populated than the top right and bottom left quadrants.  This figure also shows a difference between the shapes of the distributions of changes with size greater than and less than 100~G.  Note also the sharp drop-off at 100~G in the first panel of Figure~\ref{fchangehist}.  We find that weak field changes, $|dB_l| < 100$~G, are nearly twice as likely to decrease the field as increase it (line 4 in Table~\ref{fieldchangetable}).  This bias is greater near disk-center, where increases are more than twice as numerous as decreases, than near the limb (compare lines 8 and 10 in Table~\ref{fieldchangetable}).  On the other hand, strong field changes, $|dB_l| > 100$~G, are slightly more likely to increase the field (lines 5, 9, and 11 in Table~\ref{fieldchangetable}).  So the split in decreasing/increasing fields is dominated by weak field changes and is strongest near disk-center.  

In Table 4, we also present the Pearson correlation coefficient, $r_0$, between the field change, $dB_l$, and the background field intensity, $B_l$, for all of the representative points and for different subsets of the data.  We include in Table 4 the probability, $P(|r| > |r_0|)$, that the same number of measurements of two uncorrelated variables would yield a correlation coefficient $r > r_0$.  Overall, we find no correlation between the field changes and the background field intensities (line 1 in Table~\ref{fieldchangetable}), consistent with SH05, but we do find that weak field changes, $|dB_l| < 100$~G, show a modest negative correlation with background field intensity both near disk-center and near the limb but the correlation is stronger near disk-center (lines 4, 8, and 10 in Table~\ref{fieldchangetable}).

\section{Flux changes}
\label{fluxchanges}

\begin{table}
\scriptsize
\caption{Increasing and decreasing net and unsigned magnetic fluxes.}
\begin{tabular}{lcccc}
\hline\hline
 & No. net & No. net & No. unsigned & No. unsigned \\
 & increasing & decreasing & increasing & decreasing \\
 \hline
All & 37 & 40 & 29 & 48 \\
X-class & 19 & 19 & 12 & 26 \\
M-class & 18 & 21 & 17 & 22 \\
Disk-center & 14 & 18 & 12 & 20 \\
Limb & 23 & 22 & 17 & 28 \\
Disk-center X-class & 7 & 5 & 3 & 9 \\
Disk-center M-class & 7 & 13 & 9 & 11 \\
Limb X-class & 12 & 14 & 9 & 17 \\
Limb M-class & 11 & 8 & 8 & 11 \\
\hline
\end{tabular}
\label{fluxincdectable}
\end{table}

\begin{table}
\scriptsize
\caption{Pearson correlation coefficients and confidence levels for magnetic flux changes against magnetic flux.}
\begin{tabular}{lcccc}
\hline\hline
 & Pearson c.c. & Probability & Pearson c.c. & Probability \\
 & $r_0$ & $P(|r|\ge |r_0|)$ & $r_0$ & $P(|r|\ge |r_0|)$ \\
 & net flux &  & unsigned flux &  \\
\hline
All & -0.62 & $4.5\times 10^{-10}$ & 0.70 & $8.6\times 10^{-14}$ \\
X-class & -0.66 & $2.7\times 10^{-6}$ & 0.76 & $3.8\times 10^{-9}$ \\
M-class & -0.15 & 0.36 & 0.26 & 0.11 \\
Disk-center & -0.46 & $7.4\times 10^{-3}$ & 0.66 & $2.0\times 10^{-4}$ \\
Limb & -0.68 & $7.7\times 10^{-8}$ & 0.80 & $1.1\times 10^{-12}$ \\
Disk-center X-class & -0.50 & $9.9\times 10^{-2}$ & 0.74 & $4.3\times 10^{-3}$ \\
Disk-center M-class & -0.46 & $4.0\times 10^{-2}$ & 0.43 & $5.8\times 10^{-2}$ \\
Limb X-class & -0.72 & $1.3\times 10^{-5}$ & 0.82 & $2.9\times 10^{-8}$ \\
Limb M-class & 0.063 & 0.80 & 0.29 & 0.23 \\
\hline
\end{tabular}
\label{fluxchangetable}
\end{table}

\begin{figure*}[ht]
\begin{center}
\resizebox{0.45\hsize}{!}{\includegraphics*{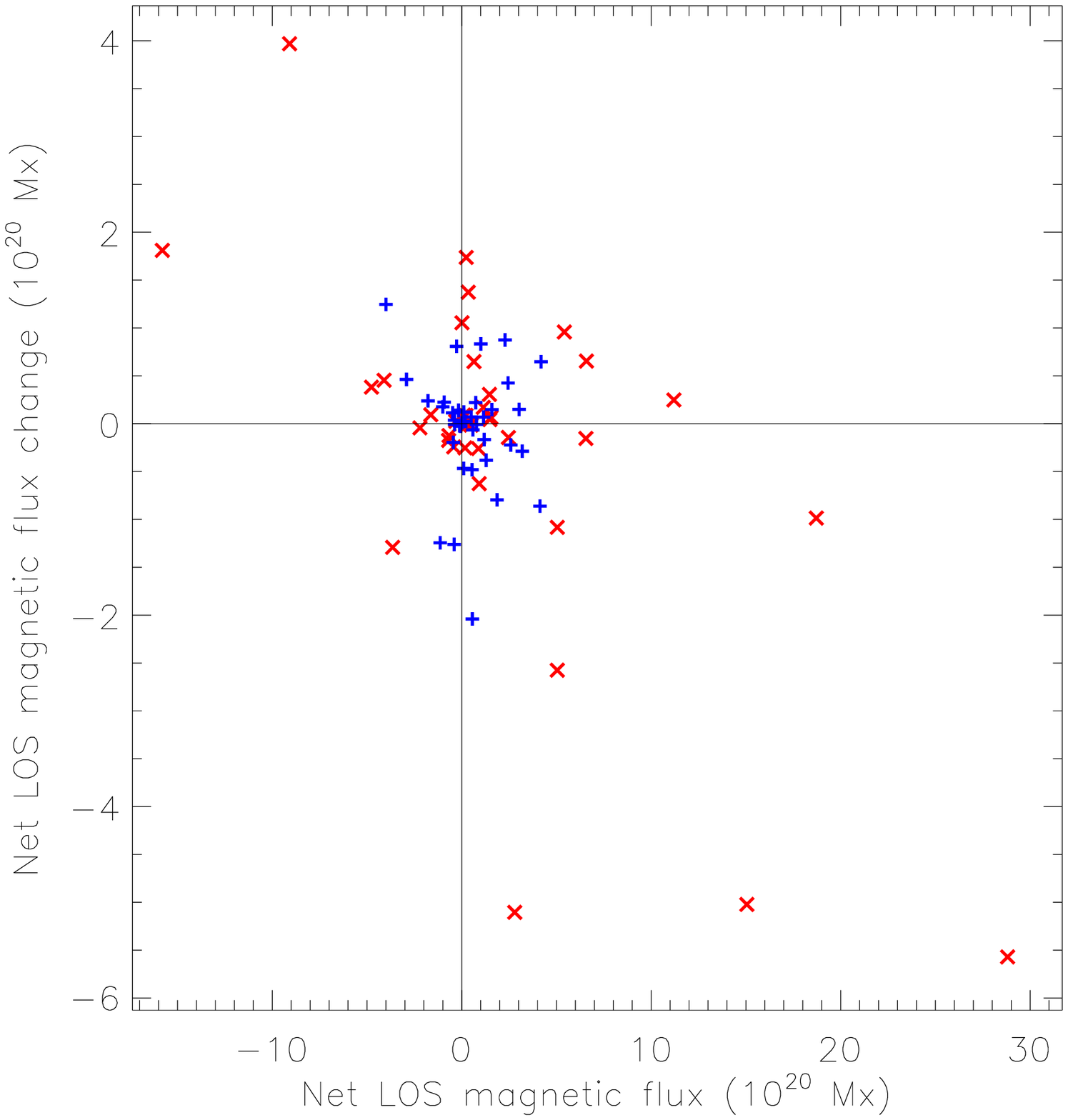}}
\resizebox{0.45\hsize}{!}{\includegraphics*{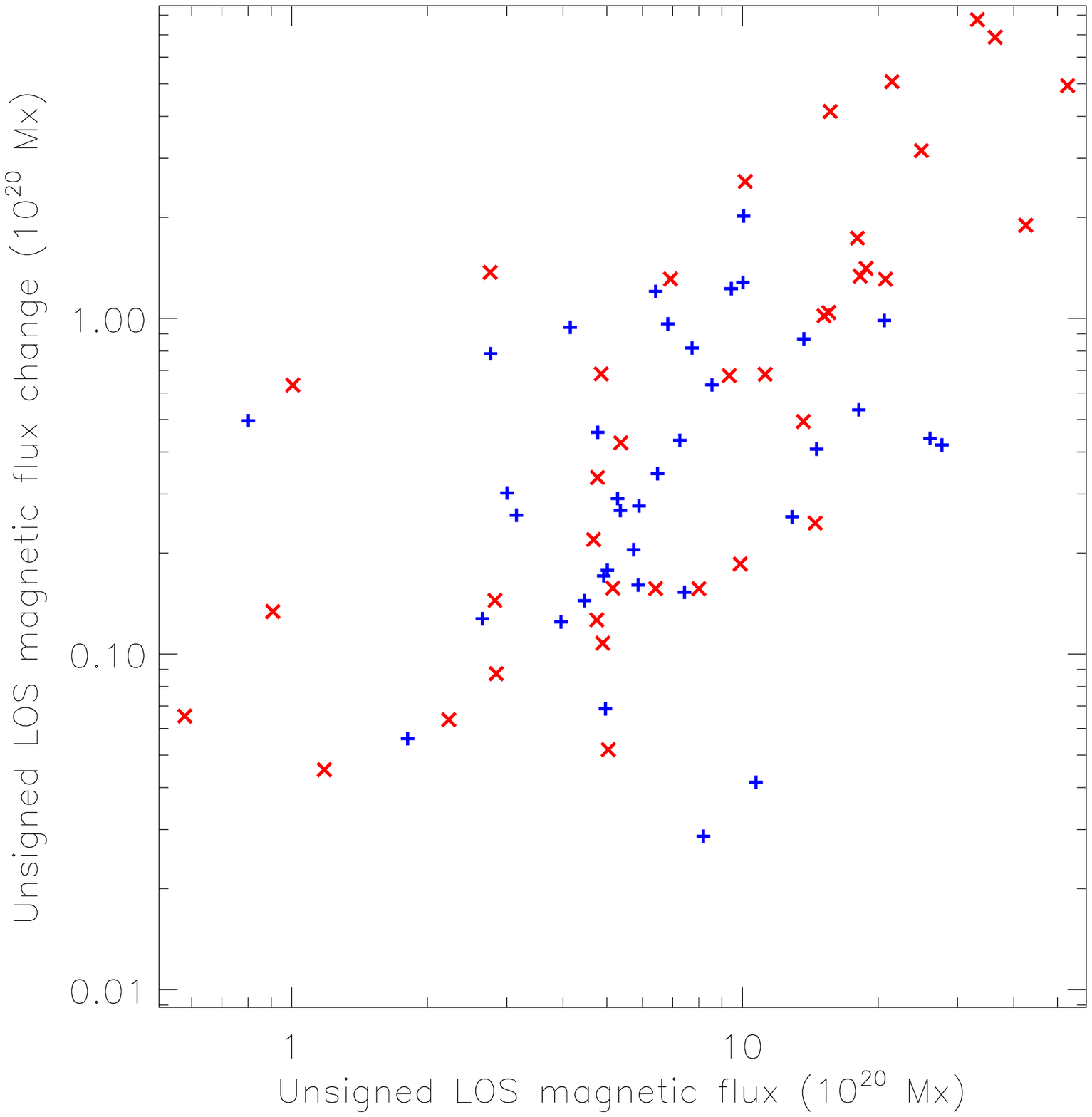}}
\end{center}
\caption{Scatter plots of the change in the net magnetic flux against the background net magnetic flux (left) and the change in the unsigned magnetic flux against the background unsigned magnetic flux (right).  Blue plus signs (+) denote M-class cases and red crosses ($\times$) X-class cases.}
\label{fluxes}
\end{figure*}

As SH05 have noted, the change in magnetic flux may be a more important physical quantity than the most significant and abrupt field change at one particular location.  The representative pixels represent the fastest and strongest field changes, free of artifacts, but do not present a complete picture of the changes to the magnetic field in an active region.  Magnetic flux calculations capture the effects of flares on entire active regions.  In this section, we estimate the change in the magnetic flux during each flare.  Because all of the pixels in a remapped image have equal area, the net flux is just the sum of the field changes over all of the pixels in the remapped image.  We eliminate from consideration, of course, those pixels for which the parameters of the fit of Equation~(\ref{atancurve}) to the time series plot do not satisfy all of the criteria (a-d) described in Section~\ref{fieldchanges}.  These flux calculations no doubt include some ``false positives'', but we expect these to average out.


Figure~\ref{fchangehist} shows histograms of the net (bottom left) and unsigned (bottom right) magnetic flux changes for all 77 flares in this study.  The vast majority of the flux changes are towards the weak end of the scale.  About half of the flares fall into the first bin in each histogram.  Like the histogram of field changes in the same figure, these histograms resemble power laws but with a stronger power index.  The flux changes have a greater range of values than the field changes do because the area is a varying parameter in the flux calculations but not in the field intensity calculations.

Table~\ref{fluxincdectable} shows the statistics for the increases and decreases in net and unsigned flux.  The net fluxes increase in approximately equal numbers in general (compare the first two columns of Table~\ref{fluxincdectable}).  Unsigned fluxes, on the other hand, decrease for nearly two thirds of the flares overall (see the top line of Table~\ref{fluxincdectable}), and for those near disk-center as well as those near the limb (lines 4 and 5 of Table~\ref{fluxincdectable}).  For X-class flares the ratio of flux decreases to increases is greater than 2:1 while for M-class flares the ratio is closer to 1:1 (compare lines 2 and 3 of Table~\ref{fluxincdectable}).  X-class flares near disk-center show three times as many decreases as increases in unsigned flux whereas X-class flares near the limb show fewer than twice as many (lines 6 and 8 of Table~\ref{fluxincdectable}.  The corresponding M-class statistics are closer to 1:1, perhaps because they are compromised by noise (lines 7 and 9 of Table~\ref{fluxincdectable}).  When we sort the data according to location, East/West or North/South or quadrants, no patterns emerge.

Wang \& Liu~(2010) found for 17 out of 18 flares occurring in $\delta$-spots that the limbward flux increased and the diskward flux decreased according to longitudinal field measurements.  It is not straightforward to compare these results with ours because the active regions in our data set do not generally have bipolar flux distributions, and those that do are not always aligned approximately East-West.

Figure~\ref{fluxes} shows the scatter plots of net and unsigned flux change against background net and unsigned flux.  The correlations between magnetic flux changes and background flux are summarized in Table~\ref{fluxchangetable}.  The statistics show a significant overall (negative) correlation between the net magnetic flux changes and the net background fluxes.  This contrasts with the overall result for field changes in line 1 of Table~\ref{fieldchangetable}.  Figure~\ref{fluxes} (left panel) also shows this overall negative correlation between net fluxes and net flux changes.  The distribution shown in this figure is very different from the corresponding distribution of field changes shown in Figure~\ref{bvsdb}.  Although only slightly more net fluxes decrease than increase, the larger net flux changes are generally decreases, hence the significant correlation between net and unsigned flux change and net and unsigned background flux (line 1 in Table~\ref{fluxchangetable}).  Furthermore, the larger flux changes are almost all associated with X-class flares, so the negative correlation is significant for X-class examples but not for M-class examples (compare lines 2 and 3 in Table~\ref{fluxchangetable}).  There is stronger correlation near the limb than near disk-center (compare lines 4 and 5 in Table~\ref{fluxchangetable}).  X-class flares show stronger correlation near the limb than near disk-center (lines 6 and 8 in Table~\ref{fluxchangetable}).  This is because the largest flux changes occur during X-class flares near the limb, a phenomenon that we will discuss further in Section~\ref{diskposition}.  On the other hand, among M-class flares disk-center cases show significant correlation with a 95\% confidence level while limb cases do not (lines 7 and 9 of Table~\ref{fluxchangetable}).  This may be because more of the M-class data near the limb are compromised by noise.

\section{Relation of field and flux changes to GOES peak X-ray flux}
\label{fieldfluxint}

\begin{table}
\scriptsize
\caption{Pearson correlation coefficients and confidence levels of field and flux changes with GOES peak X-ray flux.}
\begin{tabular}{lcccccc}
\hline\hline
 & Pearson c.c. & Probability & Pearson c.c. & Probability & Pearson c.c. & Probability \\
 & $r_{bi}$ field & $P(|r|\ge |r_{bi}|)$ & $r_{ni}$ net & $P(|r|\ge |r_{ni}|)$ & $r_{ui}$ unsigned & $P(|r|\ge |r_{ui}|)$  \\
 & intensity change &  & flux change &  & flux change &  \\
\hline
All & 0.28 & $3.3\times 10^{-4}$ & 0.38 & $5.7\times 10^{-4}$ & 0.48 & $6.8\times 10^{-6}$ \\
X-class & 0.22 & $3.7\times 10^{-2}$ & 0.32 & $5.0\times 10^{-2}$ & 0.42 & $8.1\times 10^{-3}$ \\
M-class & 0.25 & 0.14 & 0.14 & 0.40 & 0.18 & 0.27 \\
Disk-center & 0.14 & 0.27 & 0.60 & $1.8\times 10^{-4}$ & 0.37 & $3.6\times 10^{-2}$ \\
Limb & 0.29 & $4.4\times 10^{-3}$ & 0.34 & $2.2\times 10^{-2}$ & 0.49 & $5.1\times 10^{-4}$ \\
Disk-center X-class & -0.01 & 0.96 & 0.65 & $2.0\times 10^{-2}$ & 0.25 & 0.44 \\
Disk-center M-class & 0.22 & 0.21 & 0.00 & 1.00 & 0.11 & 0.65 \\
Limb X-class & 0.26 & $4.5\times 10^{-2}$ & 0.29 & 0.15 & 0.43 & $2.7\times 10^{-2}$ \\
Limb M-class & 0.12 & 0.50 & 0.38 & 0.11 & 0.39 & 0.10 \\
\hline
\end{tabular}
\label{fieldfluxinttable}
\end{table}

\begin{figure*}[ht]
\begin{center}
\resizebox{0.45\hsize}{!}{\includegraphics*{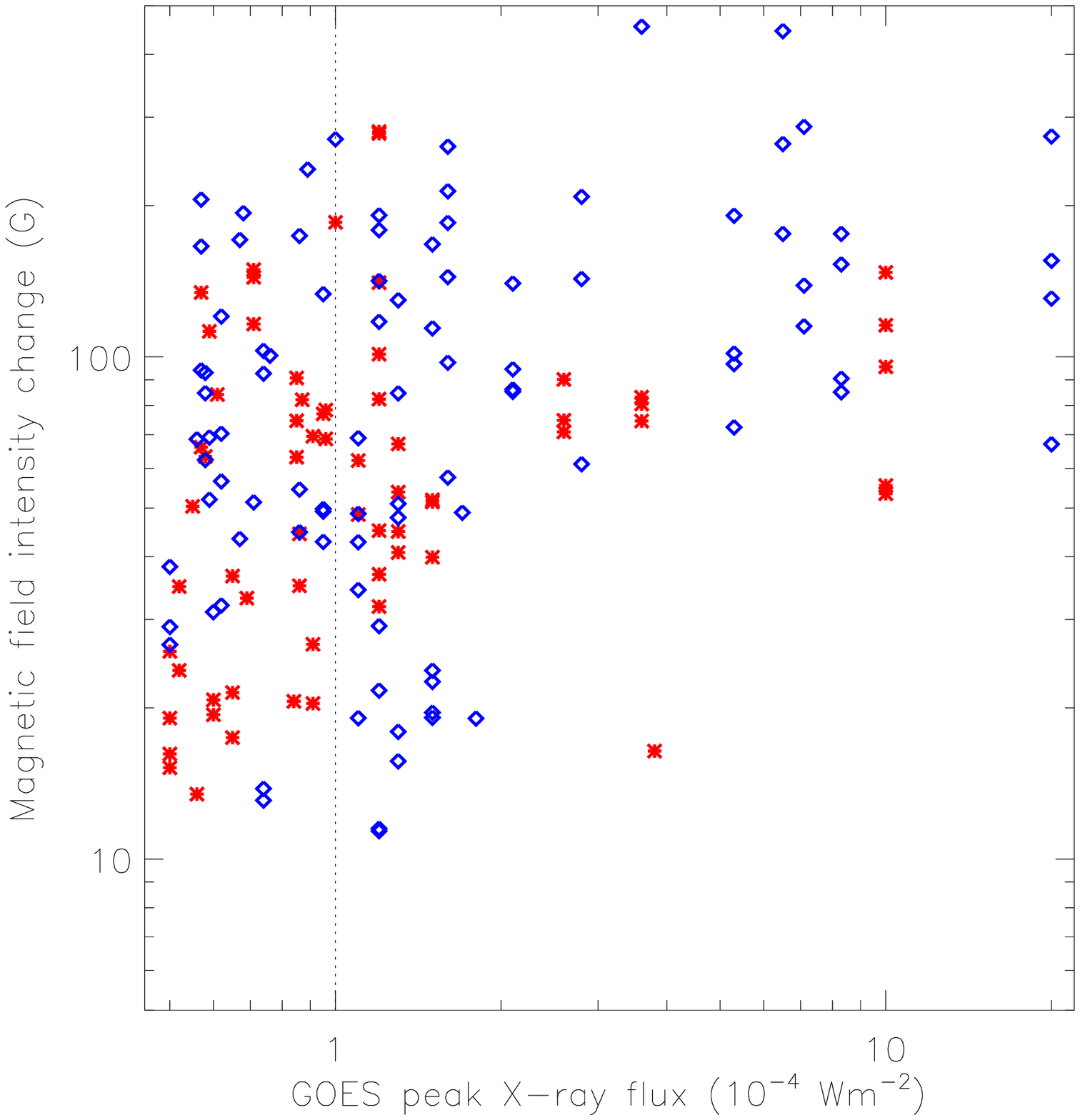}}
\resizebox{0.45\hsize}{!}{\includegraphics*{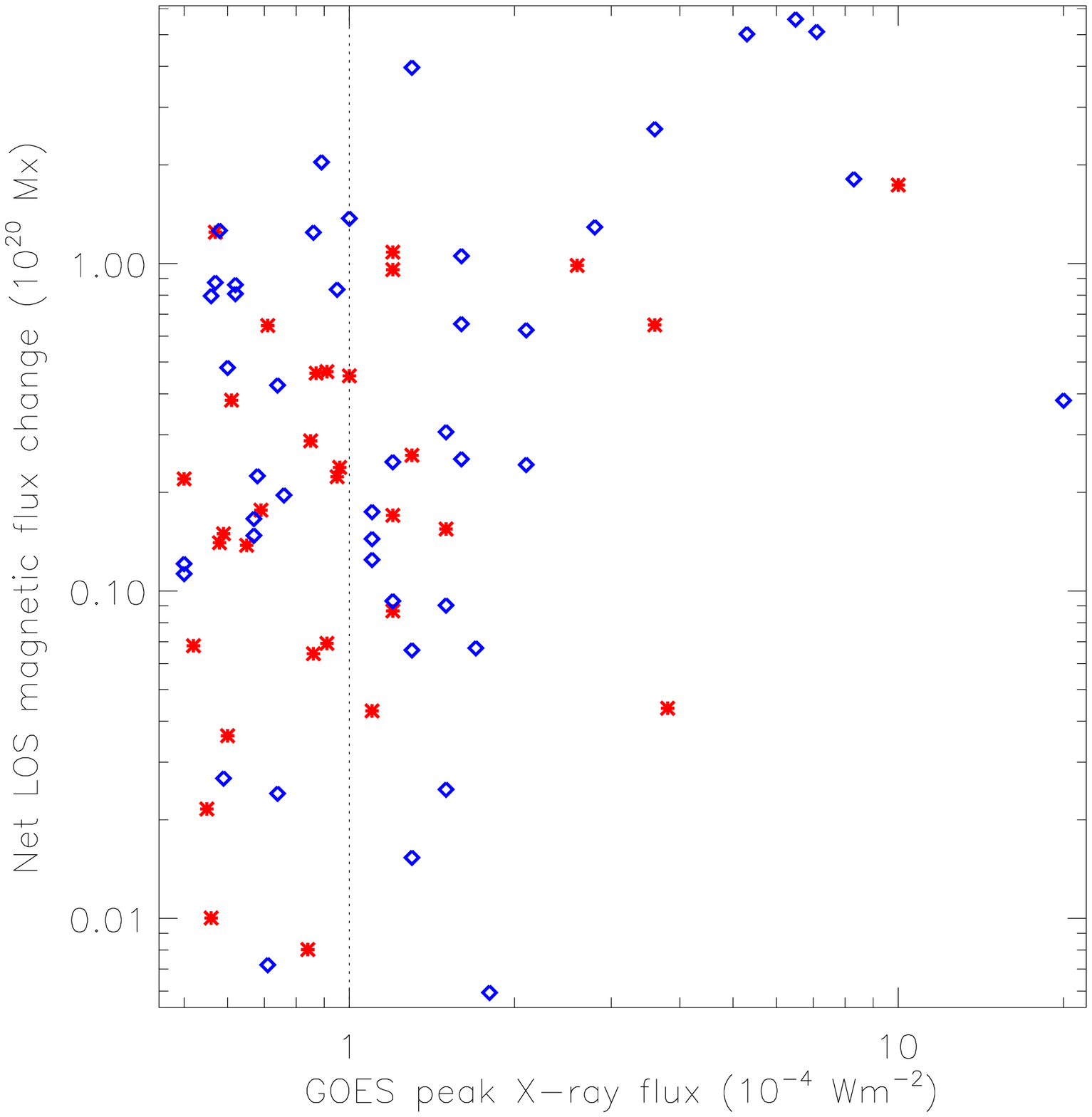}}
\resizebox{0.45\hsize}{!}{\includegraphics*{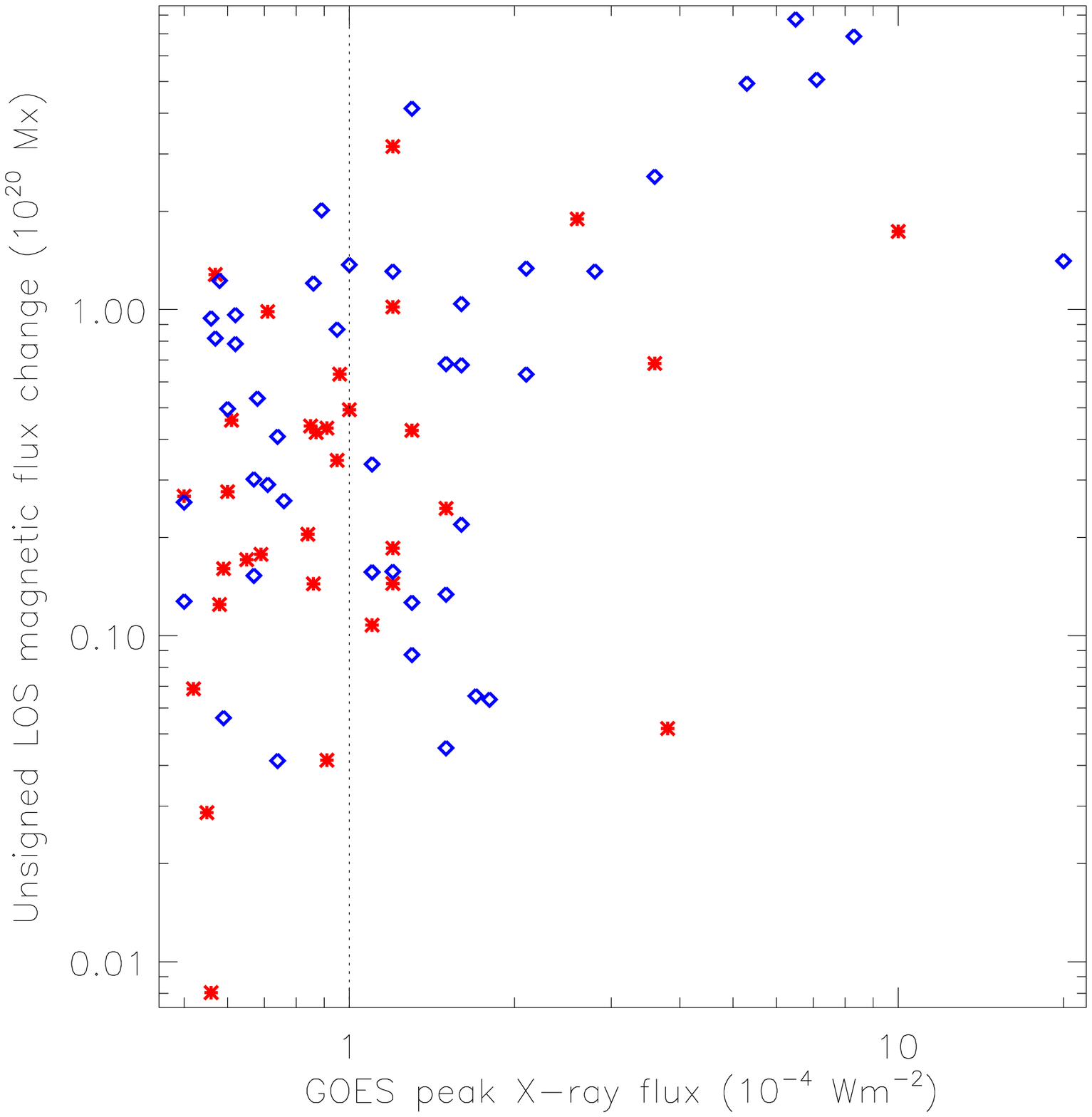}}
\end{center}
\caption{Scatter plots of magnetic field change (top left), change in net magnetic flux (top right) and change in unsigned magnetic flux (bottom) against GOES peak X-ray flux.  GOES peak X-ray flux values $< 1$ (left of dotted line) correspond to M-class flares while values $\ge 1$ (dotted line and above) correspond to X-class flares.  Red asterisks (\protect\raisebox{-.6ex}{*}) denote flares with radial position $r \le r_s/2$ on the solar disk, and blue diamonds ($\diamond$) denote cases with $r > r_s/2$, where $r_s$ is the solar disk radius in the image plane.}
\label{intfarrays}
\end{figure*}

We have seen that correlations of changes in field intensity or flux with their background values are different for GOES X-class and M-class flares and for flares observed near the limb and near disk-center.  Photospheric magnetic field properties have often been explored in the past as possible predictors of flare activity.  For example, the estimated unsigned radial magnetic flux of the active region and the unsigned flux near strong-field polarity inversion lines are two magnetic quantities that have been found to correlate with GOES X-ray flare flux (Leka et al.~2007, Schrijver~2007, Welsch et al.~2009).  In this section we seek correlations between the detected longitudinal magnetic field and flux changes and GOES X-ray flare flux.

Figure~\ref{intfarrays} shows scatter plots of the magnetic field intensity changes and the net and unsigned magnetic flux changes against the GOES peak X-ray flux.  Table~\ref{fieldfluxinttable} summarizes the correlations between the field and flux change and the GOES peak X-ray flux.  Overall, the field change, the net flux change, and the unsigned flux change all show some weak to moderate correlation with GOES X-ray flux (line 1 of Table~\ref{fieldfluxinttable}).  The correlation between field change and GOES X-ray flux is dominated by X-class flares at the limb (compare the first two columns of lines 2, 5, and 8 in Table~\ref{fieldfluxinttable}).  As discussed in Section~\ref{fieldchanges}, the median value of the field change is higher for X-class flares than for M-class flares and for limb flares than disk-center flares (Table~\ref{minmaxtable}).  All this adds up to the conclusion that X-class limb flares are slightly more likely than other flares to produce strong longitudinal field changes.  

Some correlation between the change in the magnetic flux and the GOES X-ray flux is expected, as discussed in Section~\ref{fluxchanges}, and the correlation is dominated by the X-class flares (compare lines 1 and 2 in Table~\ref{fieldfluxinttable}).  The statistically significant correlation between the net magnetic flux change and the GOES X-ray flux at disk-center is dominated by the disk-center X-class flares (compare the middle two columns of lines 4 and 6 in Table~\ref{fieldfluxinttable}).  The statistically significant correlation between the unsigned magnetic flux change and the GOES X-ray flux near the limb is dominated by the limb X-class flares (compare the last two columns of lines 5 and 8 in Table~\ref{fieldfluxinttable}), so the X-class flare theme repeats itself, but we do see two deviations from this theme.  The correlation between the net flux change at the limb and the GOES X-ray flux (middle two columns of line 5) appears to be split amongst X- and M-class flares (lines 8 and 9).  The correlation between the unsigned magnetic flux change and GOES X-ray flux near disk-center (last two columns of line 4) appears to be significant in spite of the fact that the separate M-class and X-class correlations are not. This is because these distributions occupy disjoint parts of the parameter space.  For limb flares the overall correlation (line 5) is stronger than the separate X-class and M-class correlations (lines 8 and 9) for the same reason.

We see a more significant correlation between net flux change and GOES 
X-ray flux at disk center than at the limb (but both correlations have a 
confidence level better than 5\%) and a more significant correlation 
between unsigned flux change and GOES X-ray flux at the limb than at 
disk center (but again both correlations have a confidence level better than 
5\%).  A large net flux change implies a large asymmetry between the change in 
positive flux and the change in negative flux. If the net flux change is large then positive and negative 
fluxes cannot both have moved towards or away from zero by approximately 
the same amount.  Therefore the fact that net flux changes correlate well with peak GOES X-ray 
flux near disk-center suggests a connection between X-ray flux emission and 
asymmetric vertical flux changes.

\section{Relation of field and flux changes to their position on the solar disk}
\label{diskposition}

\begin{figure*}[ht]
\begin{center}
\resizebox{0.45\hsize}{!}{\includegraphics*{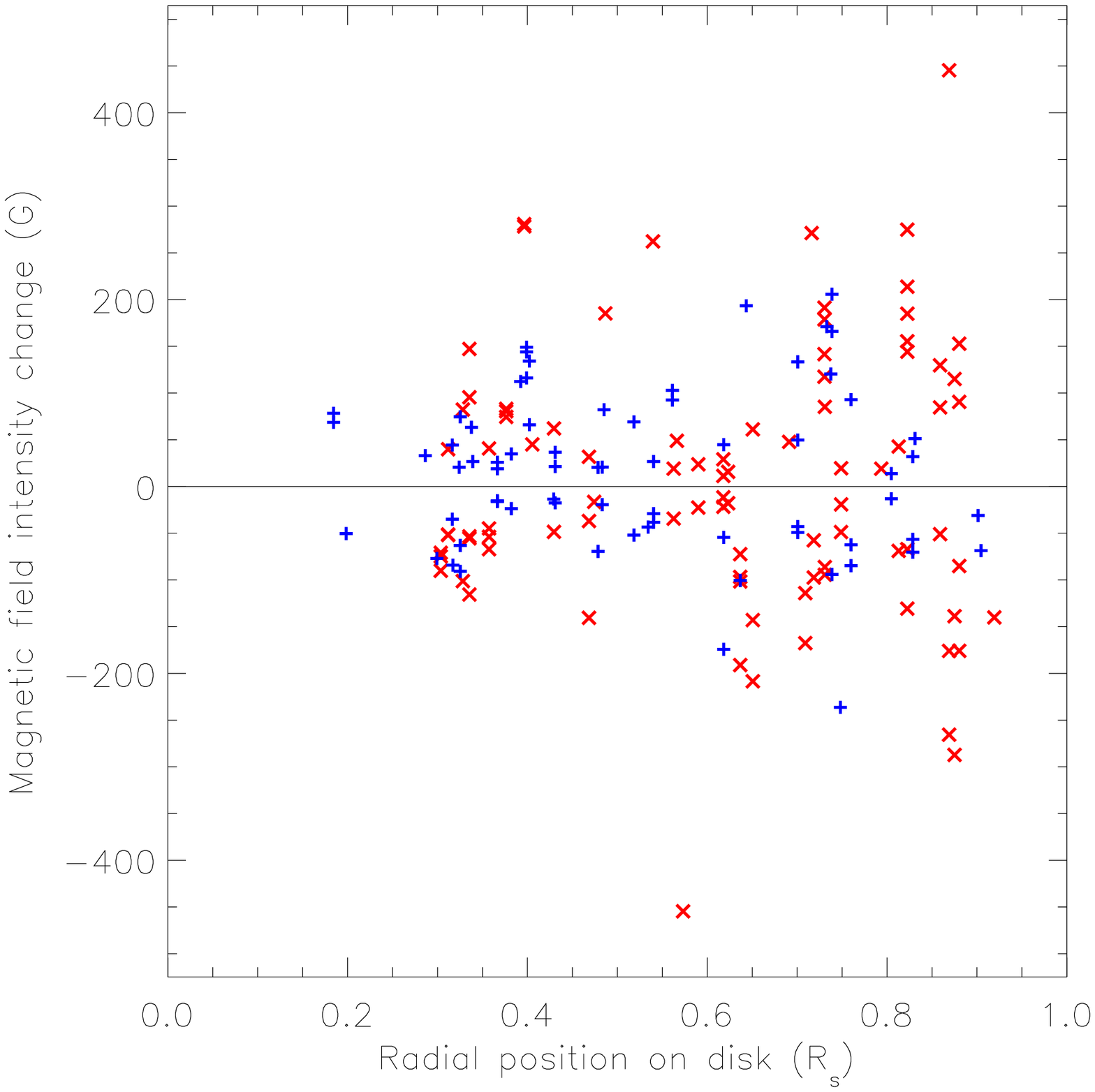}}
\resizebox{0.45\hsize}{!}{\includegraphics*{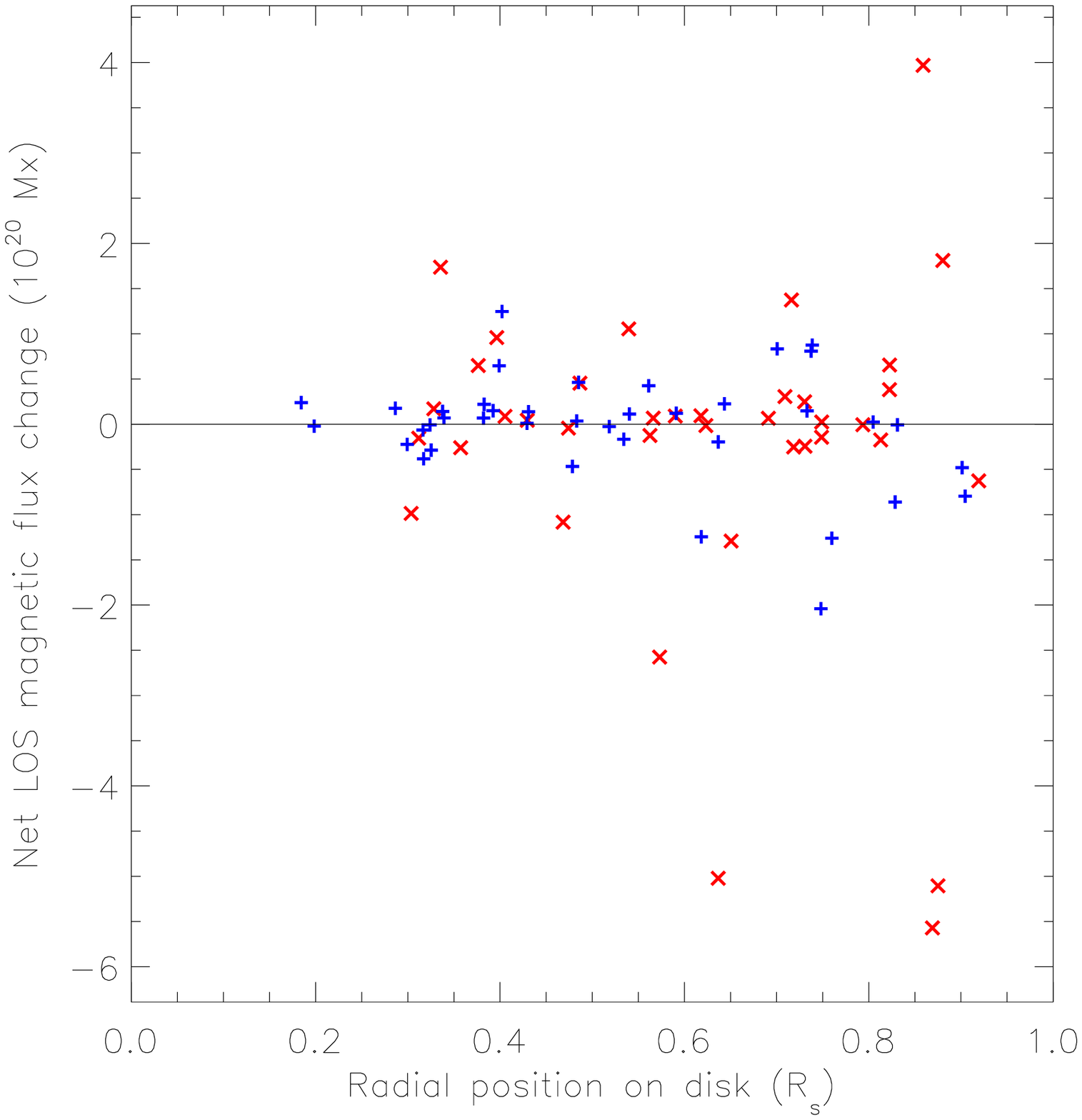}}
\resizebox{0.45\hsize}{!}{\includegraphics*{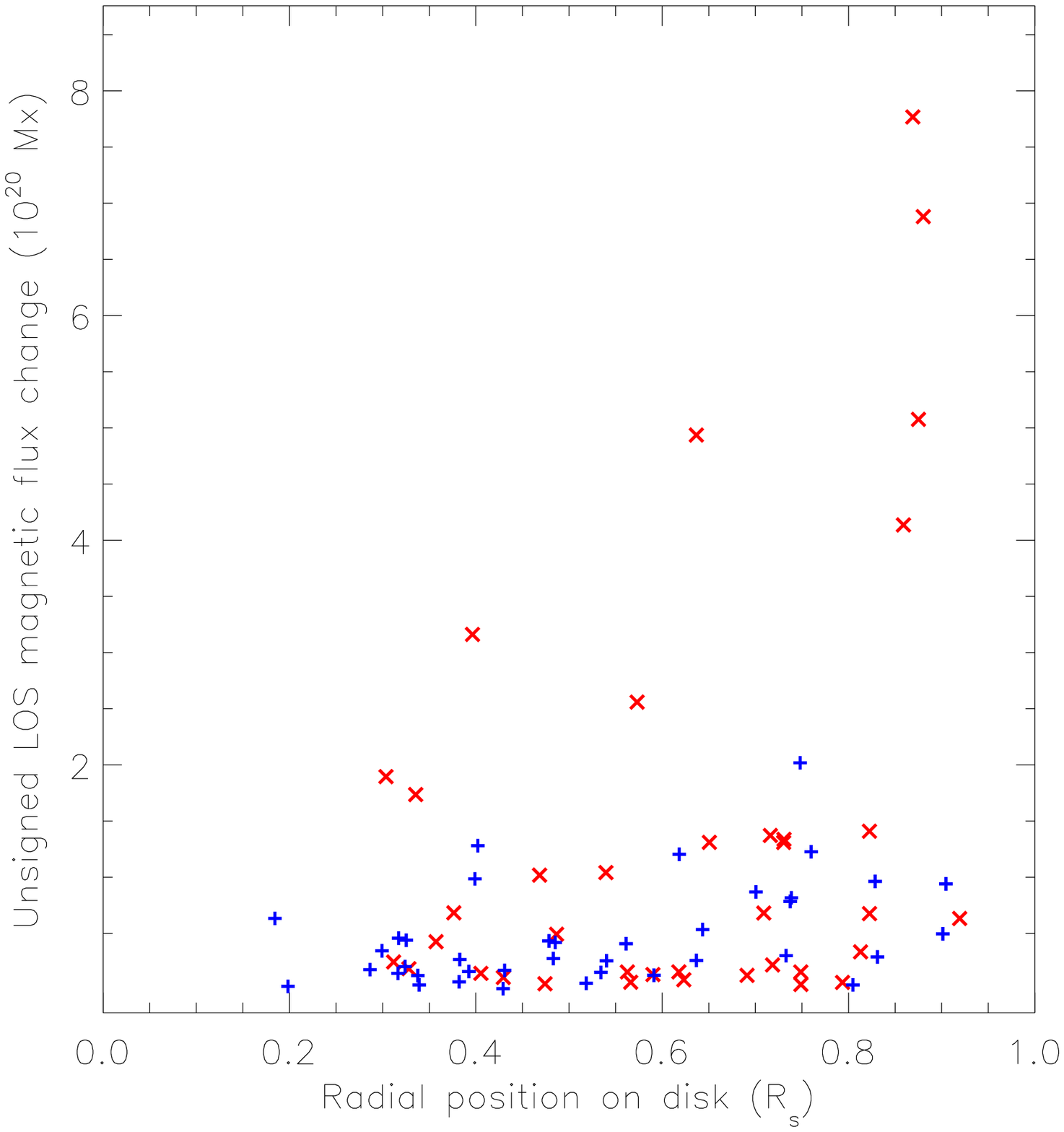}}
\end{center}
\caption{Scatter plots of the change in magnetic field intensity (top left) and the net (top right) and unsigned (bottom) magnetic flux changes against radial position on the solar disk.  Blue plus signs (+) denote M-class cases and red crosses '$\times$' X-class cases.}
\label{rhovsc}
\end{figure*}

SH05 reported that they found no trends when they sorted their data by hemisphere or by distance from disk-center.  While we do not see significant differences between the hemispheres in our larger data set, we have seen in Section~\ref{fieldchanges} that the median field change near the limb is greater than the median value near disk-center.  In this section we investigate directly the relationship between the distance from disk-center and the size of changes in magnetic field intensity and flux.

Figure~\ref{rhovsc} shows scatter plots of changes in magnetic field intensity (top left), net flux (top right) and unsigned flux (bottom) against radial position on the solar disk.  The top left plot clearly shows that strong field changes occur preferentially closer to the limb than to disk-center.  Indeed, when we sort and bin the field changes in terms of radial position on the disk in three equally populated bins, the averages and standard deviations of these bins are $32.4\pm 126.1$~G, $-18.1\pm 168.2$~G and $25.0\pm 206.7$~G.  The distributions therefore become wider with increasing radial distance from disk-center.  A larger range of longitudinal field changes is found near the limb than near disk-center.  If we exclude the two strongest field changes, one of which falls into the middle bin and the other into the limbward bin, the averages remain small and the standard deviations still increase towards the limb.  It is clear that the strongest longitudinal field changes occur close to the limb.  Therefore, the strongest changes occur in cases whose observed field component is nearly horizontal.

Net flux changes show similar spatial dependence.  The top right picture of Figure~\ref{rhovsc} shows that only rather weak net flux changes are found close to disk center while the range of flux changes increases with increasing distance from disk center.  The strongest net flux changes are concentrated close to the limb.  One interpretation of this pattern is that there is simply more longitudinal flux changing near the limb than near disk-center because most of this flux is nearly horizontal.  This interpretation is supported by the spatial distribution of unsigned flux changes.  According to the bottom picture in Figure~\ref{rhovsc}, the unsigned longitudinal flux changes by large quantities near the limb but not near disk-center.  All told, the data suggest that the photospheric fields that undergo the greatest change as a result of flares are nearly horizontal.  Such structures might include low-lying loops across neutral lines or in sunspot penumbrae.

\clearpage

\begin{figure*}[hb]
\begin{center}
\resizebox{0.45\hsize}{!}{\includegraphics*{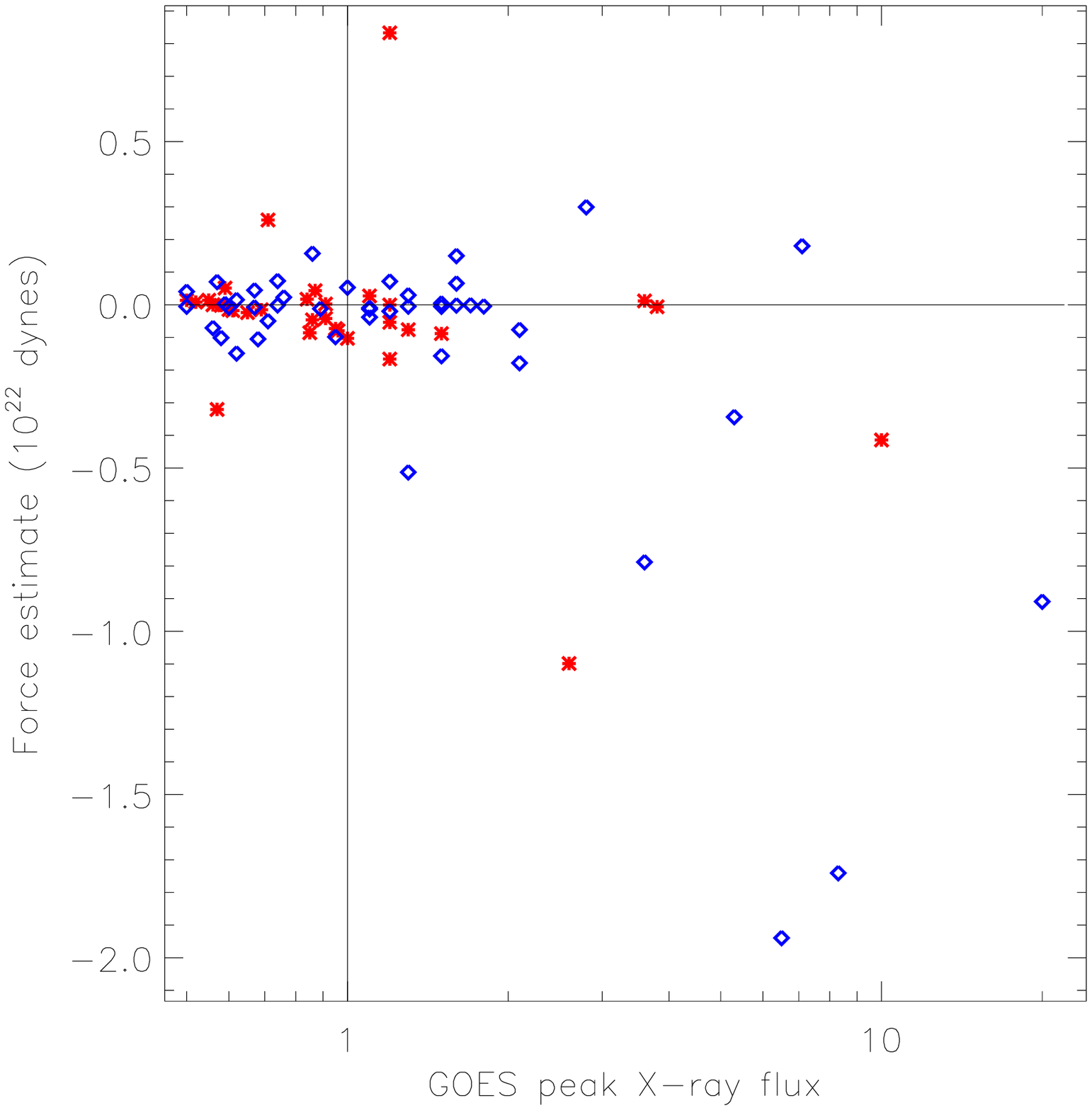}}
\end{center}
\caption{Scatter plot of Lorentz force estimates against GOES peak X-ray flux. Red asterisks (\protect\raisebox{-.6ex}{*}) denote flares with radial position $r \le r_s/2$ on the solar disk, and blue diamonds ($\diamond$) cases with $r > r_s/2$, where $r_s$ is the solar disk radius in the image plane.}
\label{forcesfig}
\end{figure*}

\begin{figure*}[ht]
\begin{center}
\resizebox{0.49\hsize}{!}{\includegraphics*{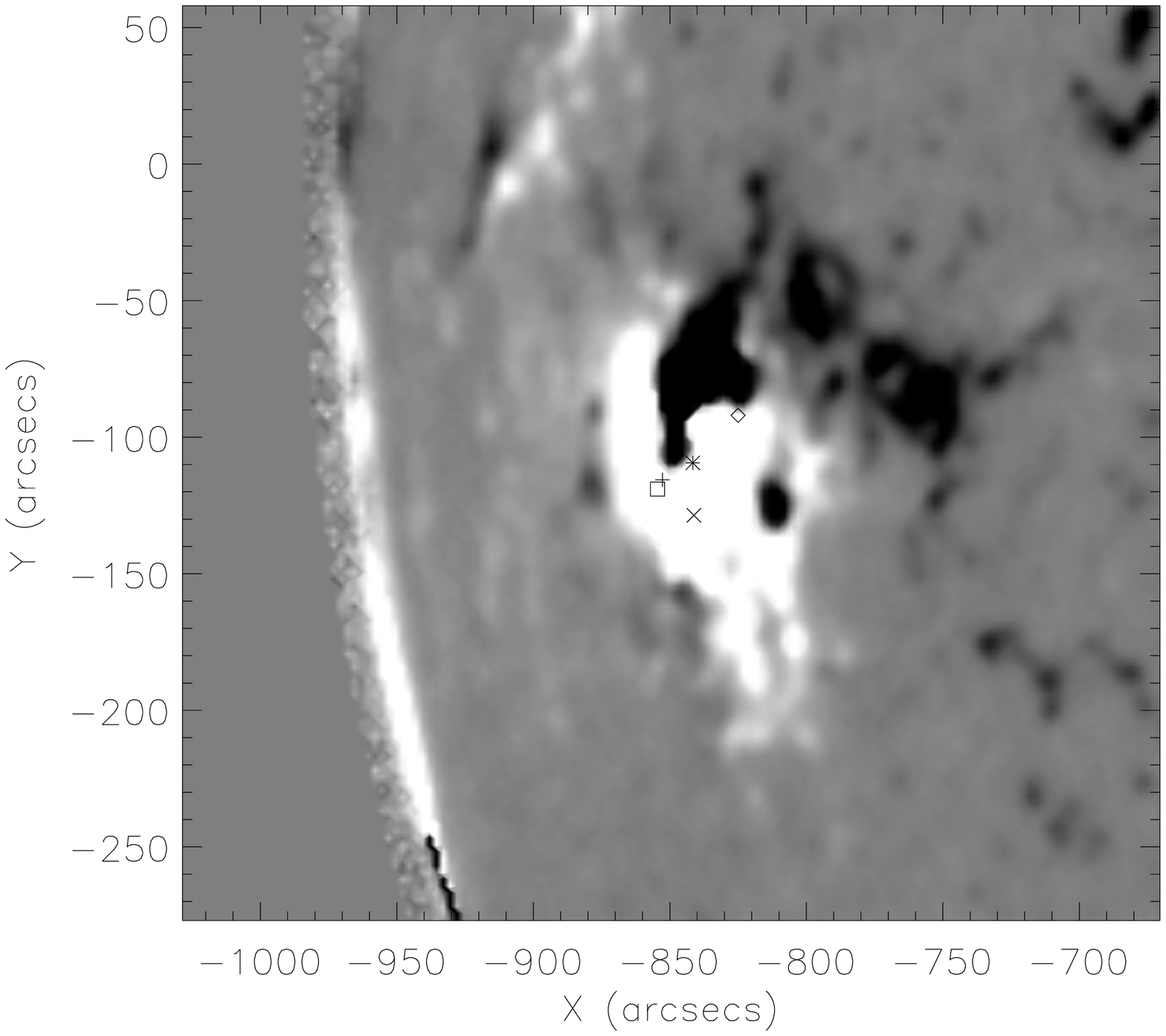}}
\resizebox{0.49\hsize}{!}{\includegraphics*{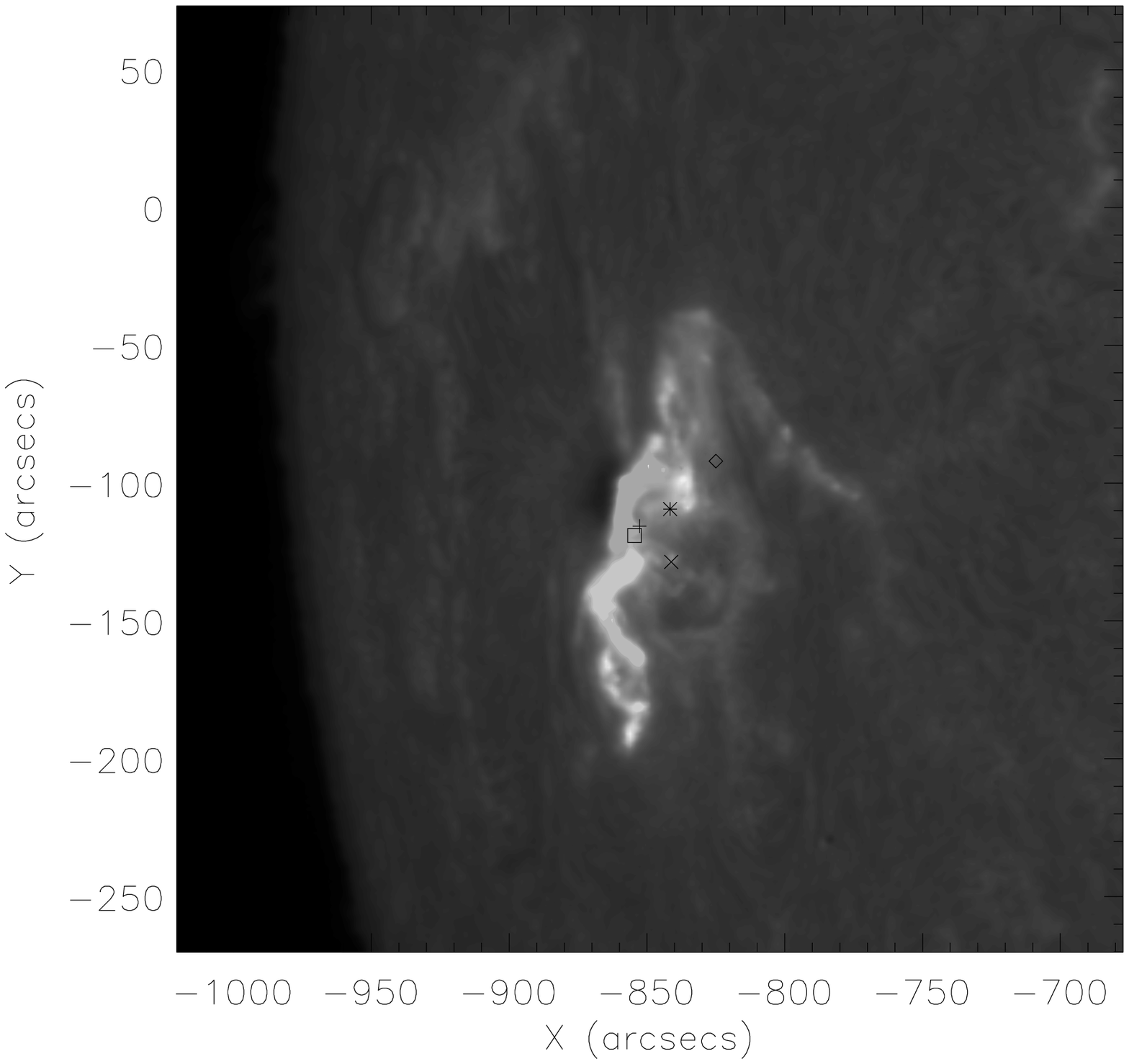}}
\end{center}
\caption{Left: a blow-up of the full-disk image of Figure~\ref{fulldisk} showing AR 10930 and the centroids of selected flare signatures.  The square ($\square$) represents the centroids of RHESSI X-ray (6-12~keV and 100-300~keV) and ISOON white-light emission, from Balasubramaniam et al.~(2010).  The asterisk (*) marks the centroid of magnetic force changes between 18:30 and 18:40~UT.  The plus (+) and cross ($\times$) symbols the centroids of the two contiguous regions of force changes between 18:40 and 18:44~UT.  The diamond ($\diamond$) marks the radiant point of the associated Moreton wave studied by Balasubramaniam et al.~(2010).  Right: ISOON H-$\alpha$ (red minus blue) image of the region at 18:43~UT with the same centroids marked.}
\label{centroids}
\end{figure*}

\begin{figure*}[ht]
\begin{center}
\resizebox{0.75\hsize}{!}{\includegraphics*{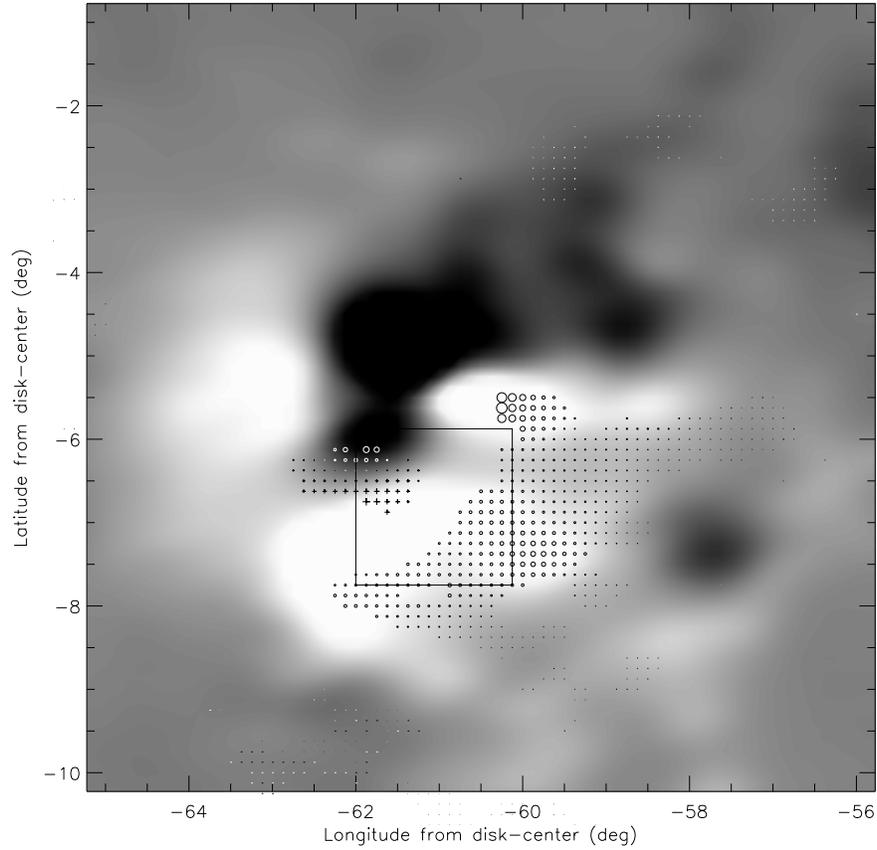}}
\end{center}
\caption{Spatial distribution of estimated force changes between 1830~UT and 1850~UT for the 2006 December 6 X6.5 flare.  Plus signs (+) denote changes where the longitudinal field increased and circles ($\circ$) changes where the longitudinal field decreased.  The symbol size represents the force change.  The forces are overplotted on the remapped magnetogram of Figure~\ref{remap}.    The black square represents the field of view of the mosaic plot in Figure~\ref{mosaic}.}
\label{forcesbala}
\end{figure*}

\begin{figure*}[ht]
\begin{center}
\resizebox{0.75\hsize}{!}{\includegraphics*{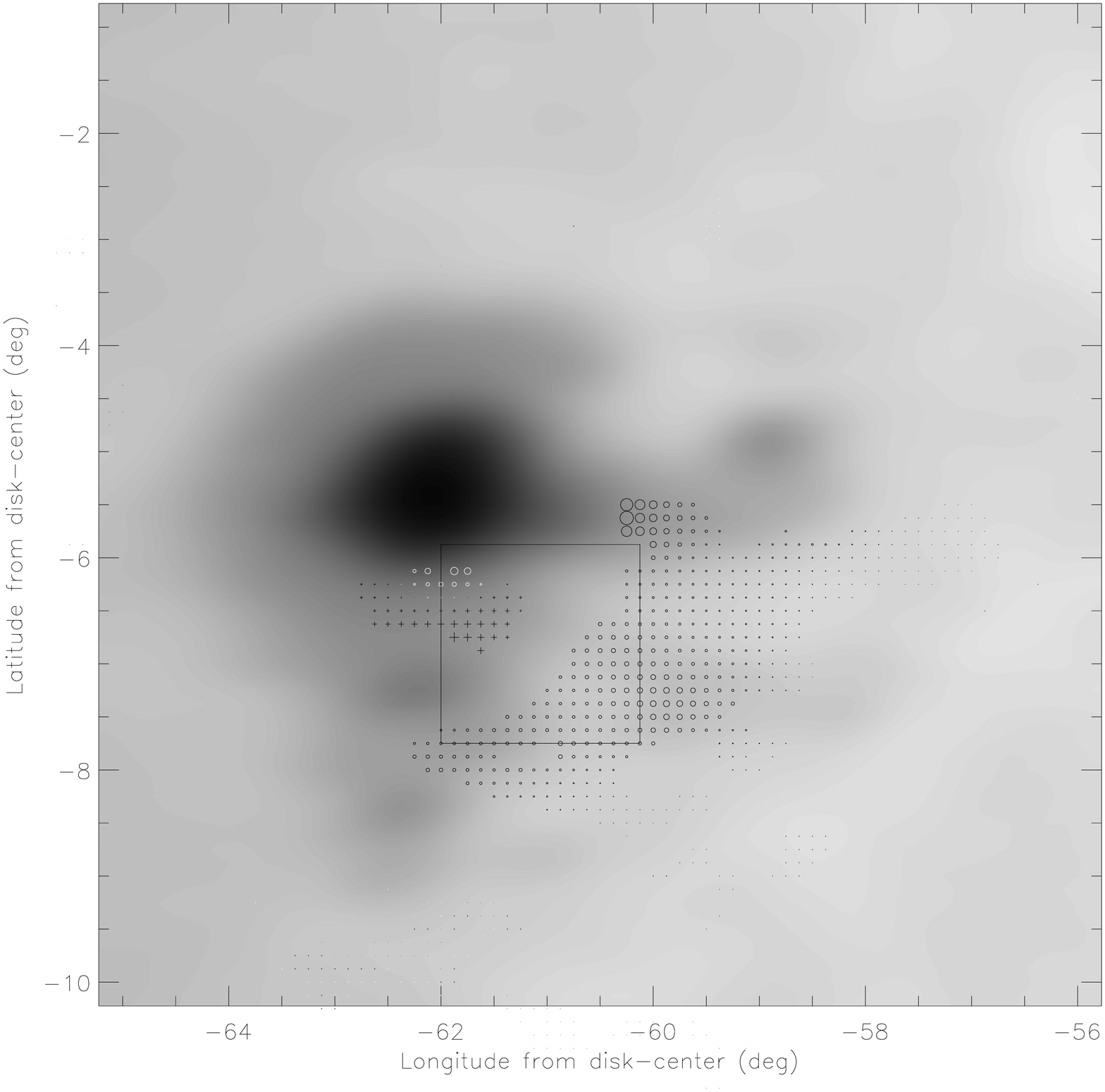}}
\end{center}
\caption{Spatial distribution of estimated force changes between 1830~UT and 1850~UT for the 2006 December 6 X6.5 flare overplotted on the remapped GONG continuum intensity image.  Plus signs (+) denote changes where the longitudinal field increased and circles ($\circ$) changes where the longitudinal field decreased.  The symbol size represents the force change.  The black square represents the field of view of the mosaic plot in Figure~\ref{mosaic}.}
\label{forcesint}
\end{figure*}

\section{Forces and energetics of the field changes}
\label{forces}

In this section we estimate the Lorentz forces applied to the photosphere by the coronal field using the photospheric magnetic field measurements.  Following a pioneering calculation by A. N. McClymont (Anwar et al.~1993, Hudson et al.~2008, Fisher et al.~2010) we can estimate how much of the released flare energy goes into reorganizing the photospheric field.  Assuming that the photosphere was in force-balanced equilibrium before the flare, a known vector field change of $(\delta B_x, \delta B_y, \delta B_z)$ results in a force imbalance whose vertical component would be

\begin{equation}
\delta f_z = ( B_z\delta B_z - B_x\delta B_x - B_y\delta B_y ) / 4\pi .
\label{zforce}
\end{equation}

\noindent The total vertical force on the photosphere could then be found by integrating $\delta f_z$ over the area over which field changes occur.  Fisher et al.~(2010) show that this expression should give a robust and accurate estimate if integrated over regions of strong field in a vector magnetogram and if the field changes are small compared to the initial field values.  Since here we only have longitudinal measurements we estimate the size of force change by

\begin{equation}
\delta f_l = B_l \delta B_l / 4\pi .
\label{losforce}
\end{equation}

Figure~\ref{forcesfig} shows a scatter plot of these force estimates against the GOES peak X-ray fluxes.  The width of the distribution increases as a function of
GOES peak X-ray flux.  The largest forces correspond
to major X-class flares while no M-class flare produces a
force stronger than about $3\times 10^{21}$ dynes.  Only one term (out of three) of the
expression for the Lorentz force in Equation~(\ref{losforce}) is available from
longitudinal data, so the sign of the force is not
determined by a single field component.  On the other
hand, we can say from the sign of Equation~(\ref{losforce}) whether
$B_l$ increased or decreased in strength.  There are 27 positive
force values and 50 negative, meaning that more of the
fields decreased than increased.  As for the largest forces of size greater than
$5\times 10^{21}$ dynes,
there are six associated with decreasing fields and one associated with an increasing field.  The largest
forces in the
sample are associated with decreasing fields.  The fact that $B_l
\delta B_l$ is more likely to be negative than
positive is
consistent with the results of Sections~\ref{fieldchanges} and \ref{fluxchanges}.  Confirmation with a full force estimate
must
await a sizable sample of
good, high-cadence vector data.  Hudson et al.~(2008) estimated that forces of size $10^{22}$~dynes can be important for the physics of seismic waves.  Several examples in Figure~\ref{forcesfig} have force budgets of this size.  Given that the forces calculated here involve only those pixels well modeled by Equation~(\ref{atancurve}), some of these forces might be significantly underestimated.  Recently, Wang \& Liu~(2010) applied Equation~(\ref{zforce}) to a BBSO vector data set for the 2002 July 26 M8.7 flare and found a vertical force change of $1.6\times 10^{22}$ dynes.  Our estimated force change for the 2002 July 26 M8.7 flare using Equation~(\ref{losforce}) is $4.3\times 10^{21}$~dynes.  The difference between the force change estimates from GONG longitudinal data and BBSO vector data is perhaps mostly due to the inclusion of the transverse field in Wang \& Liu's calculation.  We note, however, that Mart\'inez-Oliveros \& Donea~(2009) did not find good spatial correspondence between locations of abrupt, significant field changes and seismic sources in the two flares that they studied

The most impressive estimated force budget in our data set is for the 2006 December 6 X6.5 flare featured in Figures~\ref{remap}-\ref{mosaic}.  This flare has been associated with a Moreton wave, studied in detail by Balasubramaniam et al.~(2010) using H$\alpha$ images from the Improved Solar Observing Optical Network (ISOON).  This Moreton wave traveled from its source at approximately S06E63 at about 800~km/s with an azimuthal span of about 270$^{\circ}$.  A blow-up of the East-limb portion of the full-disk magnetogram of Figure~\ref{fulldisk} is plotted in Figure~\ref{centroids}, showing the flaring active region, AR 10930.  Also plotted is an ISOON H-$\alpha$ image taken at the height of the flare showing the flare emission.  As Figure~\ref{centroids} shows, the estimated central source was rather distant, 35-75~Mm distant according to Balasubramaniam et al.'s estimate, from the centroid of RHESSI X-ray and ISOON white light emission.  

This flare was also well observed by the Cerro Tololo station of the GONG network.  The data set for this flare has the largest detected number of pixels where large, clean, permanent stepwise field changes were detected of any of the 77 flares studied here, as well as the most impressive force budget.  The total forces involved amount to about $2\times 10^{22}$~dynes.  The spatial distribution of the changes occurring between 1830~UT and 1850~UT is shown in Figure~\ref{forcesbala}.  The forces are organized in two regions, a small region of relatively weak forces close to the neutral line mostly directed towards the observer (positive) and a more extended region to the south and west of somewhat larger forces directed towards the Sun (negative).  (Note that in this example the line of sight is tilted at $60^{\circ}$ with respect to the local vertical.)  While the group of positive forces is close to the centroid of the X-ray emission, the largest forces are clustered to the west, closer to the location of the focal point of the Moreton wave (see Figure~\ref{centroids}).  The centroids in Figure~\ref{centroids} show that the changes began near the negative-polarity sunspot and then propagated to the South and West.  The field changes associated with the largest forces are located in a region of strong positive magnetic flux a degree or so West of the sunspot.  The field changes themselves, at about 270~G, are significantly weaker than the strongest changes during the flare, of about 450~G, that are to be found in the region above the left part of the red line in Figure~\ref{mosaic}, but their force estimates are larger because they occur in a much stronger field.

Shown in Figure~\ref{forcesint} is a simultaneous 10-minute-averaged GONG continuum-intensity image, remapped to the same local Cartesian coordinates with the same forces plotted as in Figure~\ref{forcesbala}.  This intensity image shows the sunspot structure.  Most of the field and force changes appear to fall within the South-Western quadrant of the sunspot penumbra.  The inner penumbra has a mixture of field increases and decreases, including the strongest field changes observed during this flare (see Figure~\ref{mosaic}).  The very large contiguous region of field decreases to the South-West follows the outer penumbra, including a relatively intense outer penumbral structure due East of the sunspot where the largest force changes occur.

The 2006 December 6 X6.5 flare exemplifies many of the features characteristic of our data set.  The majority of the pixels show longitudinal field decreases while the strongest changes, those nearest the sunspot, are a mixture of longitudinal field increases and decreases.  The largest forces are associated with longitudinal field decreases, suggesting a downward collapse.  These features are typical for flares both near disk-center and, as in this case, near the limb.

The temporal distribution of the inferred longitudinal Lorentz force changes peaked between 1840 and 1845~UT, around the time of the fast acceleration phase of the associated CME (Balasubramaniam et al.~2010).  Fletcher \& Hudson~(2008) give a physical argument relating flare Alfv\'en waves with permanent photospheric field changes.  To our knowledge, no analogous argument for a CME bow shock has been given, but Fisher et al.~(2010) argue from Newton's third law that a change in the photospheric field to a more horizontal direction implies an inward impulse towards the solar interior accompanied by an equal and opposite outward force on the solar atmosphere.  We have been unable to determine whether or not a seismic wave also occurred during the flare because of the low quality of Doppler images so far from disk-center.

\section{Discussion}
\label{discussion}

\begin{figure*}[ht]
\begin{center}
\resizebox{0.80\hsize}{!}{\rotatebox{90}{\includegraphics*{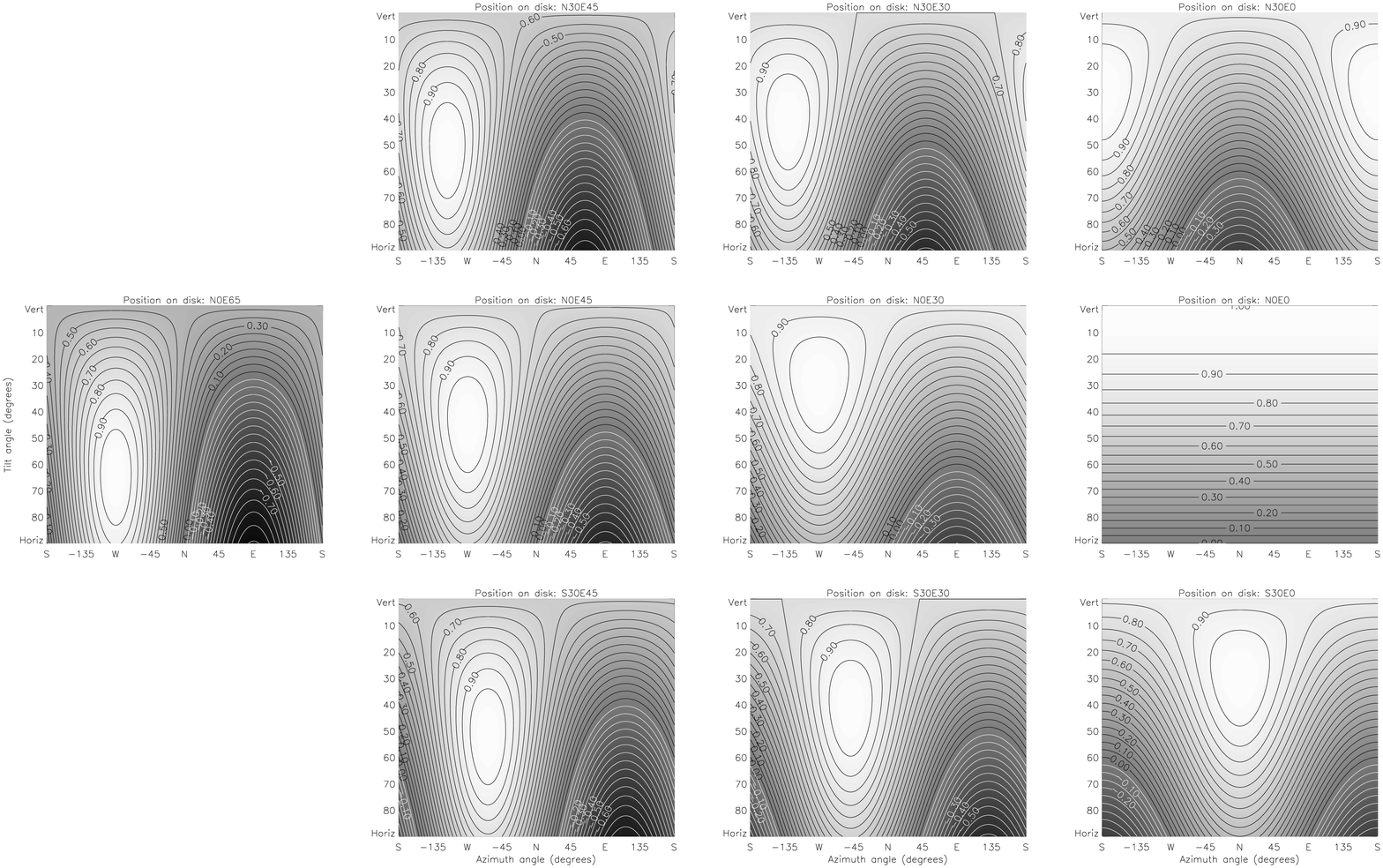}}}
\end{center}
\caption{Contour maps of the longitudinal magnetic field component $B_l$ of a unit vector as a function of the local tilt and azimuth angles.  Maps corresponding to different positions on the East half of the solar disk, indicated by the titles, are shown.  See the text for details.}
\label{forwardmod}
\end{figure*}

The observations presented in this paper provide information only on the component of the magnetic field along the observer's line of sight.  The measured net longitudinal flux generally changes during a flare at a great rate ($\approx 10^{18}$~Mx/s) and so it seems most likely that the changes in longitudinal field are caused by changes in field direction and not strength (SH05).  Proof of this can only come from precise, unambiguous, high-cadence magnetic vector field observations.  As we discussed in Section~\ref{introduction}, Fletcher \& Hudson~(2008) have shown that large-scale Alfv\'{e}n waves might transport enough energy from the flare site rapidly through the corona to change the magnetic field irreversibly at the photospheric level.  Related theoretical work (Hudson~2000, Hudson, Fisher, \& Welsch~2008, Fisher et al.~2010) predicts that the flaring photospheric magnetic fields undergo an implosion and become more horizontal as a result of flares.  In this section we assume that the change in the longitudinal field is caused entirely by a change in the direction of the magnetic vector and not by a change in its strength determine whether our observations are consistent with magnetic field vectors becoming more horizontal.

The longitudinal component $B_l$ of the magnetic field is related to the Cartesian components of the field, $B_x$, $B_y$ and $B_z$ in local heliographic coordinates and the heliographic latitudinal and longitudinal displacements from disk-center, $B$ and $L$, by the equation (e.g. Hagyard~1987),

\begin{equation}
B_l = B_z \cos{B}\cos{L} - B_y\sin{B}\cos{L} - B_x\sin{L} .
\label{loseq}
\end{equation}

\noindent The right-handed coordinates $x$, $y$ and $z$ are defined so that $z$ is normal to the solar surface and $y$ is tangent to the great circle passing through this point and the North pole.  In Figure~\ref{forwardmod} we present plots of the longitudinal field as a function of azimuthal angle and tilt angle at representative points on the East half of the solar disk.  Plots for the West half of the disk are mirror images in azimuth of the plots in Figure~\ref{forwardmod}.  Here azimuthal angle is measured in degrees clockwise from North and tilt angle ranges from $0^{\circ}$ (vertical) to $90^{\circ}$ (horizontal).  The E0N30 plot is different from the E30N0 plot by a 90-degree shift in azimuth, and the E0S30 plot by a 180-degree shift in azimuth.

At disk-center the longitudinal field is a strictly decreasing function of tilt angle.  Therefore, any decrease in longitudinal field there can only be associated with an increase in tilt.  In every other plot the picture is more complicated.  We discuss a simple example.

Suppose that the active region is a simple bipolar loop system with both polarities lying on $B=0$ (middle row of plots in Figure~\ref{forwardmod}) and that the flux in the Eastern polarity has an azimuth angle of $-90^{\circ}$ (the vector points to the West) and that the flux in the Western polarity has an azimuth angle of $+90^{\circ}$ (the vector points to the East).  Let us assume without loss of generality that the region is located in the Eastern half of the solar disk.  Then the Eastern flux as a function of tilt angle has maximum value where the tilt angle matches the angular displacement $|L|$ from disk-center.  For example, if this Eastern flux is at N0E30, N0E45 or N0E65, then its longitudinal component has maximum strength at tilt angle $30^{\circ}$, $45^{\circ}$ and $65^{\circ}$, respectively (see the longitudinal fields with $-90^{\circ}$ of azimuth in the middle row of plots in Figure~\ref{forwardmod}).  If, during a flare, the field becomes more horizontal (i.e., the tilt angle increases) according to Hudson, Fisher \& Welsch's~(2008) picture then the observed longitudinal field in the Eastern polarity would increase for tilt angles less than $30^{\circ}$, $45^{\circ}$ and $65^{\circ}$ and would decrease for tilt angles greater than $30^{\circ}$, $45^{\circ}$ and $65^{\circ}$, respectively.  On the other hand, if the azimuth angle is not exactly $-90^{\circ}$ then the longitudinal fields become more likely to decrease with increasing tilt.  For flares at locations $B\ne 0$ (see the top and bottom rows of plots in Figure~\ref{forwardmod}) the dependence of the longitudinal field on azimuth angle is offset.  At all locations, even near the limb, longitudinal field decreases can accompany tilt increases if the tilt angle is large enough.  As we have seen in Section~\ref{diskposition} many of the fields that change most during flares are likely to be nearly horizontal (i.e., they are likely to have large tilt angle).

Now we consider the Western polarity with Eastward azimuth angle ($+90^{\circ}$).  If located on $B=0$ (middle row of plots in Figure~\ref{forwardmod}) the Western longitudinal flux changes sign at a certain value of tilt.  The negative values and white contours correspond to angles where the unit vector points away from the observer and its longitudinal component is negative.  This can happen when the field vector has azimuth angle pointing away from disk-center and the field is sufficiently tilted.  For example, if the flux is at N0E30, N0E45 or N0E65 then its longitudinal component changes sign at tilt angle $60^{\circ}$, $45^{\circ}$ and $25^{\circ}$, respectively (see the longitudinal fields with $+90^{\circ}$ of azimuth in the middle row of plots in Figure~\ref{forwardmod}).  If, during a flare, the field becomes more tilted according to Hudson, Fisher \& Welsch's~(2008) picture then the observed longitudinal field strength in the Western polarity would decrease for tilt angles less than $60^{\circ}$, $45^{\circ}$ and $25^{\circ}$ and would increase for tilt angles greater than $60^{\circ}$, $45^{\circ}$ and $25^{\circ}$, respectively.

In general, the active regions in our data set are too complex for us to be able to characterize them in this way as bipolar loop systems with known azimuth and tilt angles.  Our observations only include information on the longitudinal field component.  We would need good vector observations to determine with confidence the azimuthal and tilt angles of any given field.  Without such observations, however, we can describe where in the tilt-azimuth parameter space the longitudinal field is an increasing or decreasing function of tilt.  Let us assume that the total field strength does not change significantly during the flare and that the longitudinal field change is caused only by an increase in tilt towards horizontal according to Hudson, Fisher \& Welsch's~(2008) picture.  To see where in the parameter space the longitudinal field would increase or decrease under such conditions we explore in the plots trajectories of constant azimuth angle and increasing tilt, 
i.e., trajectories of decreasing y-coordinate in the plots, and note 
whether the fields increase or decrease in strength along these trajectories.

For example, 
in the disk-center plot (right plot in middle row of Figure~\ref{forwardmod}) all longitudinal fields decrease in strength along such trajectories, as expected, whereas in the 
N0E65 plot (left plot in middle row of Figure~\ref{forwardmod}) approximately half of the longitudinal fields increase and half decrease.  Therefore, even $65^{\circ}$ away from disk-center, the longitudinal field is approximately equally likely to increase or decrease if the tilt angle increases.  Within about $65^{\circ}$ of 
disk-center, the longitudinal field decreases in most of the parameter space because of the prevalent top-heaviness of the contours, so the longitudinal field is more likely to decrease than increase if the tilt angle increases.  The 
white contours and the contours located above locations where $B_l=+1$ represent the subset of the parameter space where the longitudinal field increases if the tilt angle increases.  When a unit field vector pointing away from the observer increases in tilt at an azimuth angle pointing away from disk-center, its longitudinal component increases in strength.  Likewise when a unit vector off disk-center has tilt angle less than its heliocentric angle then an increase in tilt towards disk-center will result in an increase in longitudinal field strength.  Under all other conditions an increase in tilt, i.e., a field vector becoming more horizontal, will result in a decrease in longitudinal field strength.  Figure~\ref{forwardmod}) shows that within about $65^{\circ}$ of 
disk-center the longitudinal field is more likely to decrease than increase if the tilt angle increases as predicted by Hudson, Fisher \& Welsch's~(2008), and that this bias should be greater near disk-center than near the limb.

Is such a pattern to be found in our results?  The statistics in Section~\ref{fieldchanges} show that the observed longitudinal fields overall decreased more often than increased (94:65) and that this pattern is stronger in measurements near disk-center (42:23) than near the limb (52:42).  The pattern is more pronounced for weak ($dB_l < 100$~G) longitudinal field changes (71:37 all weak changes, 36:17 near disk-center, 35:20 near the limb) whereas the strong longitudinal field changes ($dB_l > 100$~G) do not show a statistically significant pattern.  Unsigned longitudinal magnetic flux also tended to decrease during flares both near disk-center and near the limb (Section~\ref{fluxchanges}).  The unsigned longitudinal magnetic flux decreased during $3/4$ of X-class flares near disk-center compared to nearly $2/3$ of X-class flares near the limb.  The pattern described at the end of the previous paragraph is indeed evident in our results, giving observational support to Hudson, Fisher \& Welsch's~(2008) prediction that photospheric fields become more horizontal during flares.

\section{Conclusion}
\label{conclusion}

It is now clear that the photospheric field does change during flares in general.  Extending the pioneering work of SH05 to an enlarged data set of 77 M- and X-class flares, we have reported here various separate statistics for strong/weak longitudinal field changes, X-class/M-class flares and near-disk-center/near-limb events.  As well as local longitudinal field intensity changes, we calculated changes in longitudinal magnetic fluxes.  Whereas we selected only a
few particularly clean pixels to represent field intensity changes of each flare, we used all
pixels meeting quality-control criteria to calculate magnetic flux changes.  

A summary of our results appears below.

\begin{enumerate}

\item  The median of the absolute values of the most significant and abrupt, localized changes in the longitudinal magnetic field is larger for X-class flares than M-class flares (82~G compared to 54~G) and for limb-flares than disk-center flares (85~G compared to 54~G).

\item  Overall, local, longitudinal field changes are 1.4 times more often associated with a decrease in the background field than an increase.  In more specific terms, weak field changes, $|dB_l| < 100$~G, are nearly twice as likely to decrease the field as increase it, and more than twice as likely near disk-center, whereas strong field changes, $|dB_l| > 100$~G, are slightly more likely to increase the field.


\item  Overall, we find no correlation between the field changes and the background field intensities, but we do find that weak field changes, $|dB_l| < 100$~G, show a modest negative correlation with background intensity and that this correlation is stronger near disk-center than near the limb.

\item Unsigned flux decreases for nearly two thirds of the flares overall, for those near disk-center and for those near the limb.  For X-class flares the ratio of decreases to increases is greater than 2:1.  Unsigned flux tended to decrease during flares near disk-center and near the limb.  All six net flux changes of size $2\times 10^{20}$~Mx or greater decreased the net flux.  The correlations between flux and flux changes were stronger near the limb than near disk-center because the largest flux changes occurred in X-class flares near the limb.

\item The field change, the net flux change, and the net unsigned flux change all show some modest correlation with GOES X-ray flux.  The correlation between field change and GOES X-ray flux is dominated by X-class flares at the limb.  We see a more significant correlation between net flux change and GOES 
X-ray flux at disk center than at the limb and a more significant correlation 
between unsigned flux change and GOES X-ray flux at the limb than at 
disk center.  This may connect X-ray flux emission with 
asymmetric vertical flux changes.


\item We also found a clear preference for the large changes in magnetic field intensity, net magnetic flux and unsigned magnetic flux to occur near the limb.  None of the large changes in these quantities occurred near disk-center.  Because the longitudinal fields change most near the limb and the line-of-sight direction is nearly horizontal near the limb, the data are consistent with the fields being nearly horizontal in the regions where we detect the largest field changes.  

\item We estimated Lorentz force changes using A.N. McClymont's method.  In seven X-class cases we find force changes on the order of $10^{22}$~dynes, comparable to Hudson, Fisher \& Welsch's~(2008) estimate for a force change large enough to power a subsurface seismic wave.  We also found evidence that force changes are associated more with decreases than increases in the longitudinal field, which is consistent with these forces being preferentially directed towards rather than away from the Sun.  They are therefore consistent with Hudson, Fisher \& Welsch's~(2008)'s picture of photospheric fields becoming more tilted during flares and may be important in the generation of seismic waves.

\item By considering the possible relations between actual magnetic vector tilts and the associated longitudinal field components at chosen locations on the solar disk, we found that if the field tilt only increases (the vector becomes more horizontal) during a flare, the longitudinal field can either increase or decrease, whether whether near disk-center or near the limb.  However, decreases would likely outnumber increases at all parts of the disk that we investigated (within $65^{\circ}$ of disk-center) and more so near disk-center than near the limb.  We find such patterns in our data, again consistent with Hudson, Fisher \& Welsch's~(2008)'s picture of photospheric fields becoming more horizontal during flares.

\end{enumerate}

While this work is based on a large data set, about twenty thousand magnetograms, the physical picture that emerges is incomplete.  Additional information could fill significant gaps in our understanding of the fields studied here.  Since May/June 2002 1-minute continuum intensity images have been produced by the GONG network.  Using these we can determine how strong field changes and penumbral intensity changes are related.  Umbral changes are more difficult to detect because umbral fields are very strong resulting in higher noise levels and because the dark umbral background makes flare-induced line profile transients more likely.

To investigate the relationship between the observed photospheric field changes and related changes in the corona, the magnetograms need to be supplemented with observations of higher atmospheric layers, such as H-$\alpha$ filaments and EUV loops.  Using images from NASA's Transition Region and Coronal Explorer (TRACE) satellite, SH05 found excellent spatio-temporal agreement between changes in the photospheric magnetic field and increases in brightness at foot-points of flare ribbons.  SH05 also found that the magnetic field changes appeared to cross the active regions at speeds ranging from 5-30~km/s.  Using H$\alpha$ images from Yunnan Observatory for one flare they found a strong spatio-temporal correlation between a propagating magnetic field change and the motion of an H$\alpha$ ribbon.  Further simultaneous observations of propagating field changes and flare ribbon motions might shed much light on the causes of the field changes.  Of the 15 flares that SH05 studied, they analyzed EUV data for three of them and H-$\alpha$ data for one.  These were very laborious procedures because of the differing spatial and temporal resolutions, fields of view (and vantage point in the case of TRACE) and because the H-$\alpha$ and EUV signals derive from above the photosphere.  However, this kind of work is essential if we are to understand the interactions between the photosphere and the corona.

This work has focused on the longitudinal field and flux changes without studying the changing morphology of the fields and their interactions during flares.  A feature-tracking algorithm such as YAFTA\footnote{http://solarmuri.ssl.berkeley.edu/$\sim$welsch/public/software/YAFTA} (Welsch \& Longcope~2003) can identify magnetic flux systems and trace their evolution and interaction in time.  Preliminary experiments with the magnetograms for the 2006 December 6 X6.5 flare show abrupt morphological changes corresponding to the stepwise field changes reported here.  A future study will characterize this behavior.

Finally, the interpretation of these observations of longitudinal field changes is complicated by the fact that they do not include the full field vector.  Sensitive, high-cadence vector data from the Vector Spectro-magnetograph (VSM) instrument on NSO's Synoptic Optical Long-term Investigation of the Sun (SOLIS) telescope and from the Helioseismic and Magnetic Imager (HMI) on board NASA's Solar Dynamics Observatory (SDO) spacecraft will allow us to extend this work in various ways.  Kubo et al.~(2007) studied a time series of Hinode vector data of AR10930 with 4-minute cadence covering the 2006 December 13 flare revealing many interesting field changes, but did not investigate abrupt stepwise changes in the vectors.  We can verify using VSM or HMI that the longitudinal field changes are caused by changes in direction as we expect and not strength, determine whether the field vectors become more or less tilted with respect to the vertical during flares and derive estimates of full Lorentz force vectors associated with the field changes.


\acknowledgements{We thank the referee for comments that helped us to write a more readable manuscript.  We thank Brian Harker, Jack Harvey and Frank Hill for stimulating discussions and helpful comments on the manuscript and Sean McManus for reading very many images from tape.  We also thank Pete Marenfeld for adding the color to Figure~\ref{mosaic}.  This work utilizes data obtained by the Global Oscillation Network
Group (GONG) program, managed by the National Solar Observatory, which
is operated by AURA, Inc. under a cooperative agreement with the
National Science Foundation. The data were acquired by instruments
operated by the Big Bear Solar Observatory, High Altitude Observatory,
Learmonth Solar Observatory, Udaipur Solar Observatory, Instituto de
Astrof\'{\i}sica de Canarias, and Cerro Tololo Interamerican
Observatory.}

\end{document}